\documentclass[aps,prstab,twocolumn,showpacs,superscriptaddress,groupedaddress,secnumarabic,amssymb, nobibnotes, aps, prd]{revtex4} 

\usepackage{amsmath}
\usepackage{graphicx}
\usepackage{subfig}
\usepackage{caption}
\usepackage{siunitx}
\begin{document}

\preprint{APS/123-QED}

\title{Improving Surrogate Model Accuracy for the LCLS-II Injector Frontend Using Convolutional Neural Networks and Transfer Learning}
\thanks{contact: lipigupta@uchicago.edu}%

\author{Lipi Gupta}
\affiliation{University of Chicago, Physics Department, Chicago, IL}
\author{Auralee Edelen}
\affiliation{SLAC National Accelerator Laboratory, Menlo Park, CA}
\author{Nicole Neveu}
\affiliation{SLAC National Accelerator Laboratory, Menlo Park, CA}
\author{Aashwin Mishra}
\affiliation{SLAC National Accelerator Laboratory, Menlo Park, CA}
\author{Christopher Mayes}
\affiliation{SLAC National Accelerator Laboratory, Menlo Park, CA}
\author{Young-Kee Kim}
\affiliation{University of Chicago, Physics Department, Chicago, IL}

\date{\today}

\begin{abstract}
 Machine learning models of accelerator systems (“surrogate models”) are able to provide fast, accurate predictions of accelerator physics phenomena. However, approaches to date typically do not include measured input diagnostics, such as the initial beam distributions, which are critical for accurately representing the beam evolution through the system. In addition, these inputs often vary over time, and models that can account for these changing conditions are needed. As beam time for measurements is often limited, simulations are in some cases needed to provide sufficient training data. These typically represent the designed machine before construction; however, the behavior of the installed components may be quite different due to changes over time or static differences that were not modeled. Therefore, surrogate models that can leverage both simulation and measured data successfully are needed. We introduce an approach based on convolutional neural networks that uses the drive laser distribution and scalar settings as inputs for a photoinjector system model (here, the Linac Coherent Light Source II, LCLS-II, injector frontend). The model is able to predict scalar beam parameters and the transverse beam distribution downstream, taking into account the impact of time-varying non-uniformities in the initial transverse laser distribution. We also introduce and evaluate a transfer learning procedure for adapting the  surrogate model from the simulation domain to the measurement domain, to account for differences between the two. Applying this approach to our test case results in a model that can predict test sample outputs within a mean absolute percent error of 7.6\%. This is a substantial improvement over the model trained only on simulations, which has an error of 112.7\% when applied to measured data. While we focus on the LCLS-II Injector frontend, these approaches for improving ML-based online modeling of injector systems could be easily adapted to other accelerator facilities.
\end{abstract}

\maketitle

\section{Introduction}

Physics simulations of particle accelerators are essential tools for predicting optimal settings for different running configurations (e.g. changing the bunch charge, bunch length). Injector systems are particularly difficult to model accurately \emph{a priori} because of nonlinear forces such as space charge at low beam energies. These simulations can also be computationally expensive, which can be prohibitive during the design stage as well as for online use in accelerators. In many cases, a single simulation can require minutes to hours to complete; this is a time-scale which is too long for interactive online use in the accelerator control room. This also makes it difficult to conduct systematic comparisons with measured data and account for deviation between the idealized simulation and the as-built accelerator. In addition, obtaining machine time to characterize accelerator components can be rare, especially at large facilities with high demands on beam time. 

Thus, there is a general need for fast and reliable models which can be used for online prediction, offline experiment planning, and design of new setups. Fast models would also enable more thorough investigation of differences between physics simulations and the real machine. There is also significant effort~\cite{aedsurro2,7454846, mlwhitepaper,PhysRevAccelBeams.23.044601, scheinker2020advanced,PhysRevLett.121.044801, PhysRevLett.124.124801, hanuka2019online} towards using model-based control methods in real-time machine operation and tuning, with the goal to achieve faster, higher-quality tuning. Machine learning (ML) methods may help to automate tasks such as switching between standard operating schemes, or correcting small deviations that result in poor beam quality. Fast-executing, accurate machine models can aid the development and deployment of these control methods.

Machine learning (ML) based surrogate models are one avenue toward developing fast, reliable, and realistic models of accelerators. For injector systems, data generation and model training requires significant computational resources, but once trained, ML models offer orders of magnitude faster execution speed over classical simulation methods. Amongst the many ML algorithms available, neural network (NN) based surrogate models are being widely applied for addressing the issue of execution speed and obtaining fast, non-invasive predictions of beam parameters. Several studies have verified that ML-based models can be used to support fast optimization, particularly when trained using data that spans the operational range of the physical inputs~\cite{FAST,aedsurro2,PhysRevAccelBeams.23.044601,PhysRevAccelBeams.21.112802,2019surroNeurIPS,2019surroNeurIPSKomkov,IBIC2019VD}.

While surrogate models trained on simulation are fast enough for use in online operation, the issue of how accurate these models are with respect to the real accelerator system also needs to be addressed. In many cases, large discrepancies are present due to simplifications assumed in the simulations or calibration errors. In addition, the physics simulation is often a static representation of the designed machine, and not an evolving representation of the physical machine. For example, in many photocathode injector systems, changes over time in the drive laser profile have a significant impact on the beam behavior. Typically, simulations are conducted with idealized initial beam distributions, or only a few example distributions from measurements. This can be a significant source of error between model predictions and measured data~\cite{Gulliford2015}. Overall, simulations tend to represent ideal conditions, and therefore ideal beam dynamics. Machine learning models trained on simulation data thus reproduce these discrepancies between the ideal and as-built machine behavior.

One way to work around this issue is to train surrogate models on measured data, but in many cases there is not enough data to do so, especially as the number of input settings or output measurements being sampled increases. Typical operation of an accelerator often leaves large gaps in the parameter space, and due to limited time available for dedicated machine characterization studies, it may be challenging to collect sufficient training data to produce a surrogate model that is reliable across a broad range of inputs. Similarly, injector surrogate models to date do not include the full transverse laser distribution measurements as inputs, resulting in a loss of important time-varying information for accurately predicting the beam behavior. 

The new Linac Coherent Light Source (LCLS) superconducting linac at the Stanford Linear Accelerator Center (SLAC), or LCLS-II, is one such facility that could benefit from having fast, accurate models for use in experiment planning, online prediction of beam distributions, and model-based control. Of particular importance is the injector, which sets key initial characteristics of the electron bunch, such as the overall emittance. As is the case for many injector systems, the LCLS-II injector is outfitted with a virtual cathode camera (VCC) that measures the transverse laser distribution. Changes to and non-uniformities in this laser distribution significantly impact the beam behavior.

Here, we introduce a multi-faceted, ML-based approach to address these issues by accounting for variation in the VCC image and conducting domain transfer between the simulation and measured data. First, we train a NN model based on Astra~\cite{astra} simulations of our test case, the LCLS-II injector, over a wide range of the input settings; this range is broader than is typically used for most ML-based surrogate models of accelerator systems to date. We also include changes to the initial distribution of the photocathode drive laser, as represented by measured and simulated VCC images.  We demonstrate that including the VCC image as an input to the model improves the accuracy of the surrogate model with respect to the real machine, and we show that it can accurately predict beam output for unseen (out-of-distribution) VCC images. This is essential for using the model on a real accelerator, where in many cases VCC images are likely to change from day-to-day. In this case, we combine scalar setting inputs for the injector with a Convolutional Neural Network (CNN) to do the image processing. A similar approach was taken in simulation in \cite{FAST}, and here we take the next step of including measured VCC images as inputs. Finally, we show we are able to compensate for the difference between the injector simulations and measurements by using transfer learning \cite{pratt1993discriminability, thrun2012learning}, resulting in a surrogate model that is more representative of the real machine and can interpolate between VCC images more accurately than a model trained only on measured data.

As part of this process, we also carried out a detailed study of the sensitivity of the simulated output to changes in the initial beam distribution, as seen on VCC images, to determine whether using the images directly would confer an advantage over scalar fits to the beam distribution. We also conducted a characterization study of the LCLS-II injector to compare measurements to simulations.

The main contribution from this work are as follows: (1) a characterization of the LCLS-II injector frontend, (2) introduction of an approach using measured VCC (laser distribution) images and a convolutional neural network to improve the accuracy of injector surrogate models, (3) demonstration of a  transfer learning method to pre-train the injector surrogate model in simulation and fine-tune on machine measurements. While the demonstration is specific to the LCLS-II injector, the approach can be used for other injector systems, and we have made our code publicly available to help facilitate this.

\section{Characterization Studies of the LCLS-II Injector}

In setting up a realistic surrogate model for the LCLS-II injector, it was important to assess how simulation results vary when using a realistic laser profile, as compared to an idealized Gaussian or uniform profile (as is assumed in most start-to-end injector optimizations). This determines whether it is necessary to include the full VCC image as an input to the model, or whether bulk metrics such as laser radius and an assumption of a Gaussian or uniform profile would be sufficient. We also characterized the LCLS-II injector with measured data scans and compared these to simulation data. This was done both to help improve the underlying physics simulation and assess both the need for and viability of using transfer learning for this system to account for differences between the simulation and measurement domains.

\subsection{The LCLS-II Injector}

The LCLS-II injector will produce the electron beam for the LCLS-II superconducting linac that is presently being commissioned at SLAC.

\begin{table}
\caption{Operating parameters for the LCLS-II injector. \label{tab:inj-params}}
\begin{tabular*}{\columnwidth}{@{\extracolsep{\fill}}lrl}
\textbf{Parameter}& \textbf{Value} & Unit  \\ \hline
Charges          & 100 & pC \\
Laser FWHM       & 20 & ps  \\
Laser Radius     & 1 & mm \\
Field on Cathode & 20 &  MV/m \\
Repetition Rate  & 1&  MHz 
\end{tabular*}

\end{table}

\begin{figure*}[ht!]
    \centering
    \includegraphics[width =\textwidth]{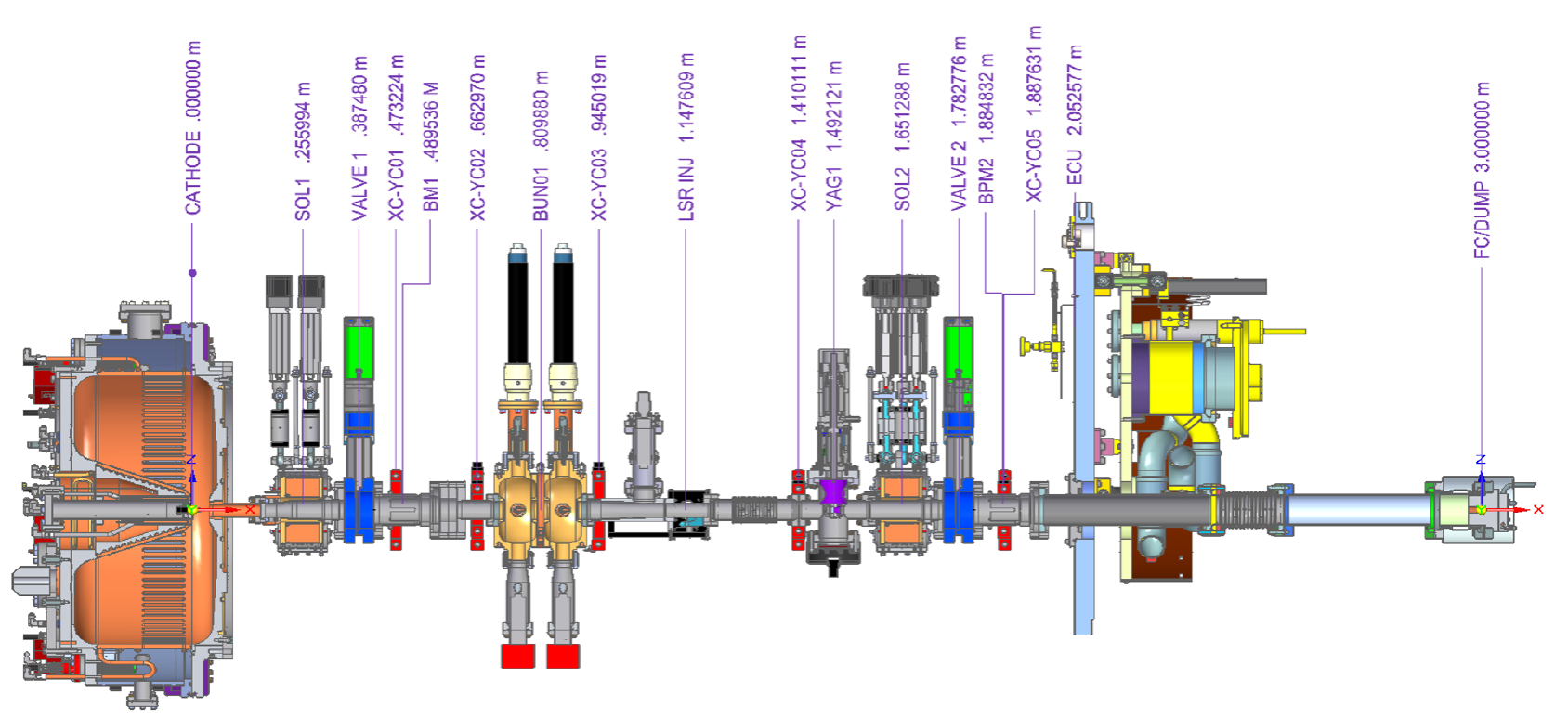}
    \caption{A schematic of the LCLS-II injector, showing each component and its position along the beamline~\cite{PhysRevSTAB.15.103501}.}
    \label{fig:inj-schematic}
\end{figure*}
The injector, shown in Fig.~\ref{fig:inj-schematic}, will be used for the LCLS-II project. The expected operating parameters for the injector are shown in Table~\ref{tab:inj-params}. At the time of experimentation, the injector was in early commissioning.

With such a high repetition rate, the cathode field gradient is lower than low-repetition rate photocathode injectors~\cite{PhysRevSTAB.15.103501}. Thus, the kinetic energy of the electrons as they are emitted and injected into the buncher is relatively low; up to 750~keV. At this energy, the dynamics are space-charged dominated. To study the dynamics in this regime and optimize the parameters for operation, particle-in-cell simulations are necessary. As such, these calculations can take several minutes to complete. 

There are several options for simulation tools, especially for accelerator injector simulations, which rely on sophisticated space charge calculations. For this study, all simulated data was generated using Astra~\cite{astra}, and particle generation was done using distgen~\cite{distgen}. The SLAC-developed Python wrapper LUME-Astra~\cite{lume-astra} was used to create, set-up, and process simulated data.

Measurements of laser input distributions and associated solenoid scans (where the electron beam size in one transverse direction was measured while the solenoid value was changed) were recorded for use in surrogate model training. These scans were taken at several different beam charges, ranging from about 1~pC to 25~pC. Machine values such as magnet currents and beam charge were also recorded. 

\subsection{Comparison Between Simulation and Measured Data}

\begin{figure*}[t]
    \centering
    \includegraphics[width = \textwidth]{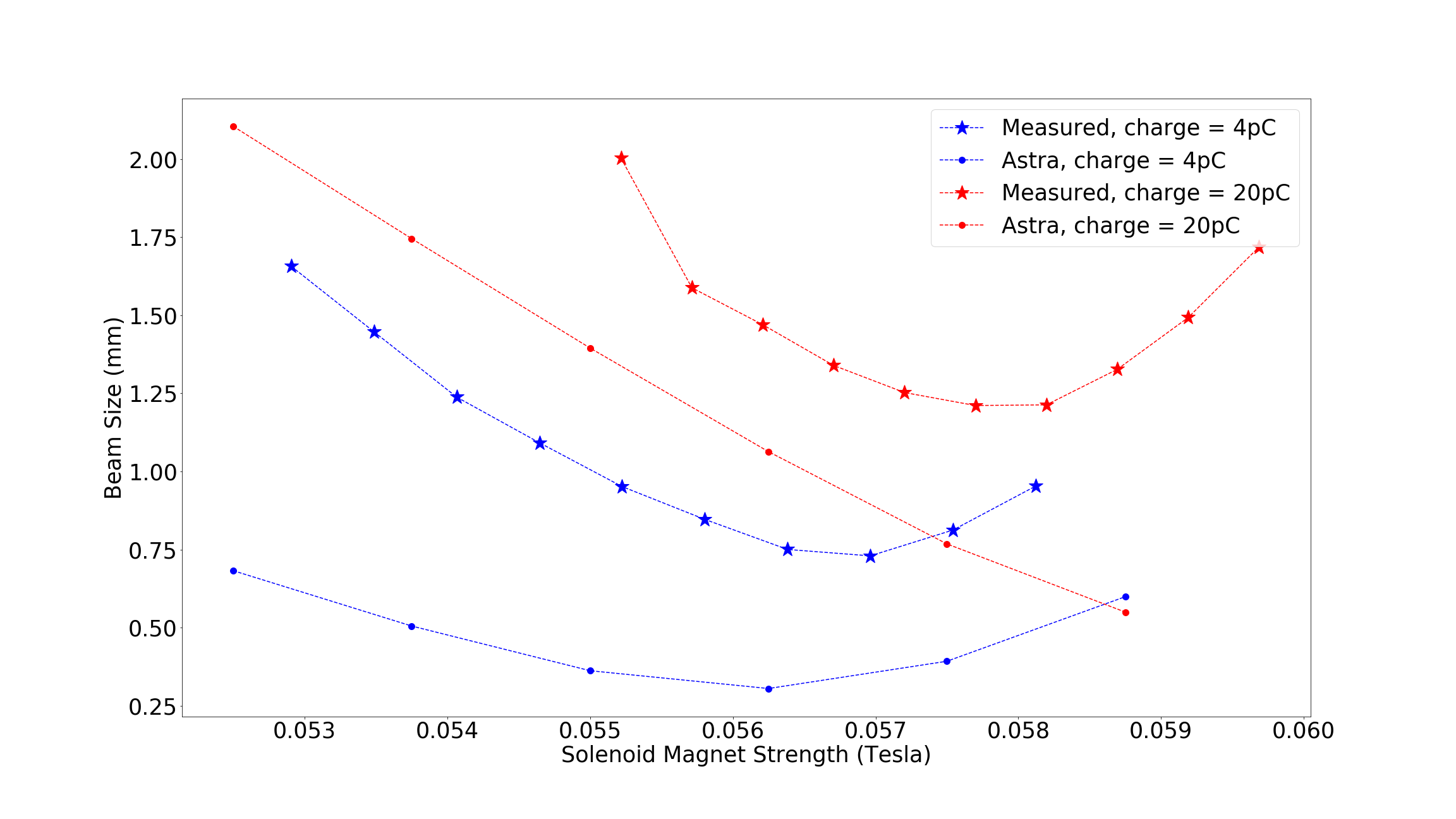}
    \caption{A comparison of simulated output values from Astra and the measurement values with the same machine parameters.}
    \label{fig:Meas to Astra}
\end{figure*}

Here we show a comparison between measured data and the associated simulated values for the LCLS-II injector. In Fig.~\ref{fig:Meas to Astra}, the beam sizes measured during two solenoid scans on the injector are compared to the predictions from Astra. The initial particle distribution for these comparison scans were generated by sampling Super Gaussian distributions in the transverse dimension, for 10,000 particles. The laser diameter was archived during the measurement and used for re-simulation. With first order matching of inputs (charge, radius, gradient, solenoid strength), there are clear discrepancies between simulation and measurement, as shown in Fig.~\ref{fig:Meas to Astra}. After investigating these errors through simulation scans, it is clear that the gun gradient and laser radius can have a large effect on the focal point of the solenoid scan. While the gradient on the cathode has a significant impact on beam size and the location of the beam waist, there was not a spectrometer located in the beam line when this data was taken. Therefore, the exact beam energy at the time of measurement is not definitively known. Indirect measurements were attempted with a small corrector, but the results were inconclusive given they returned infeasible energy values (i.e. higher energies than possible given the amount of RF power supplied to the gun). This leaves the exact gun energy to be determined, and a probable cause of discrepancy in the measurement.

\subsection{Simulation-based Sensitivity Studies}

\subsubsection{Generation of Electron Distributions from Laser Distributions}

Using the measured laser distribution to sample particles for space charge simulation can minimize discrepancies between measurement and simulated output values \cite{PhysRevSTAB.18.083401}. Therefore, in order to attempt to create more realistic simulated data, real laser profiles were used to generate particle distributions to be tracked in LUME-Astra. These laser profiles were collected at the LCLS-II. The measurement is conducted as follows: the laser beam is passed through an optical splitter, such that approximately 5\% of the beam intensity is directed towards a camera. The distance between the splitter and the camera is analogous to the distance between the splitter and the cathode, i.e. the transverse size of the laser at the camera location should be equal to the size at the cathode. The intensity at the camera is recorded, providing an image of how the laser intensity at the cathode appears.

We compared how the the beam sizes would differ when the initial particles were sampled from a realistic laser distribution and an ideal one. An idealized Super Gaussian (SG) density distribution is a common choice. The SG distribution, $\rho$, is given as function of radius $r$ of the form: 
\begin{equation}
2\pi\rho(r) = \frac{1}{\Gamma(1+\frac{1}{p})\sigma^2}{\rm{exp}}[ - (\frac{r^2}{2\sigma^2})^p]
\label{eq:supergaussian}
\end{equation}
where $\sigma$ is the standard deviation, $\Gamma$ denotes the Gamma-function, and $p$ is the SG parameter. In the limit $p \to 1$, the SG is a standard Gaussian distribution. However, as $p \to \infty$, the SG distribution approaches a flat-top distribution. By parameterizing $p$ by $p = 1/\alpha$, the $\alpha$ parameter is bounded by $[0,1]$. 

In order to make an idealized transverse SG distribution from measured VCC images, the following optimization procedure was completed. The measured laser profile was projected onto each transverse axis. A SG profile was then generated using an initial $\alpha$ value, and projected similarly into each transverse axis. An optimizer then iterated $\alpha$ via a Brent minimization algorithm~\cite{Brent}, to minimize the residual between the projections. Many of the measured laser profiles are highly non-uniform or highly irregular edges. Therefore, results for which the per pixel percent error distribution standard deviation of less than $20\%$ were selected. A nominal VCC and associated SG were chosen for the sensitivity analysis.

\begin{figure*}[ht!]
\begin{tabular}{cc}
  \includegraphics[width=65mm, trim= 0 0 70 75, clip ]{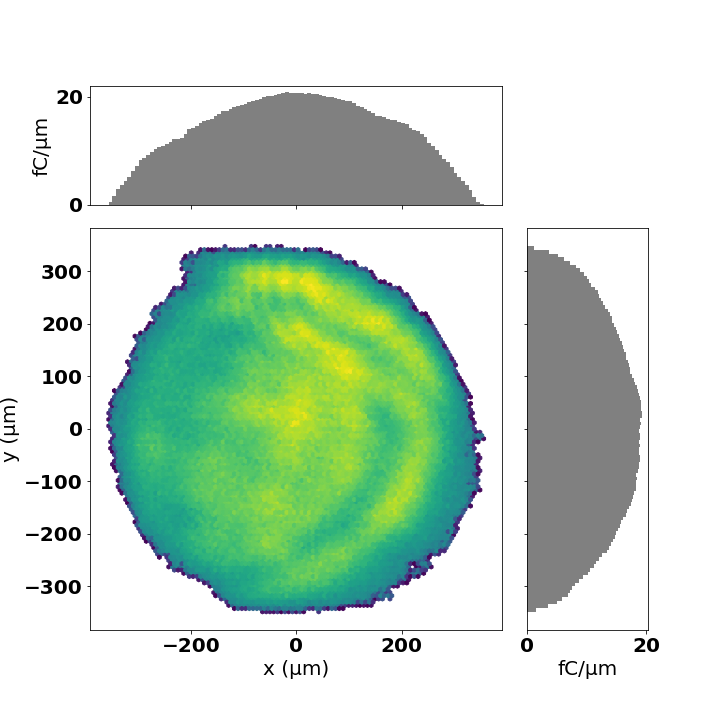}&   
  \includegraphics[width=65mm, trim= 0 0 70 75, clip]{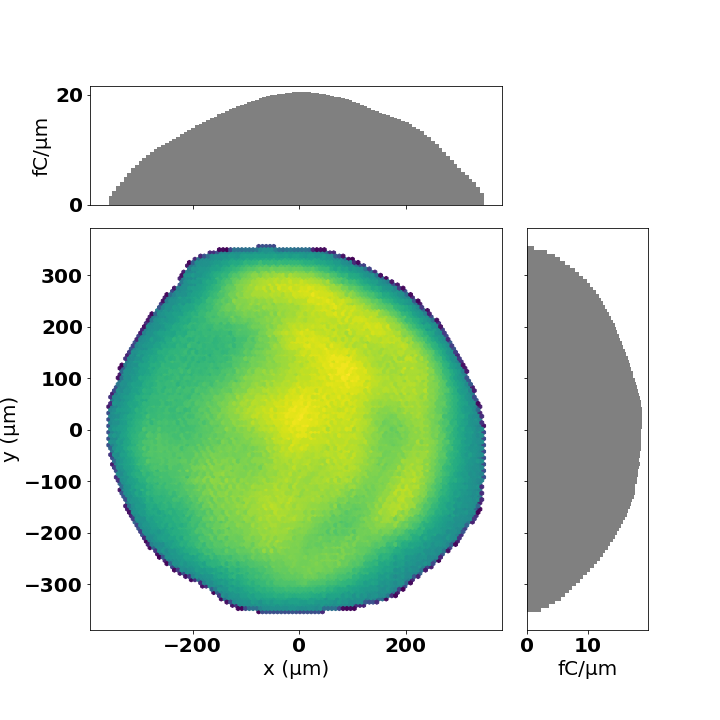} \\
  {(a) VCC-generated initial electron distribution.}&   
  {(b) Smoothed VCC-generated initial electron distribution.} \\
  
  \includegraphics[width=65mm, trim= 0 0 70 75, clip ]{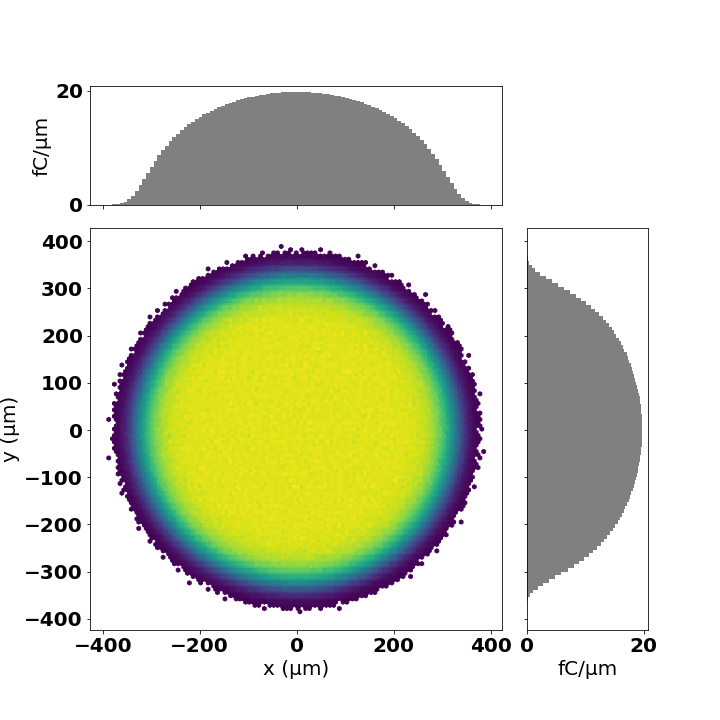}&
 \includegraphics[width=65mm, trim= 0 0 70 75, clip ]{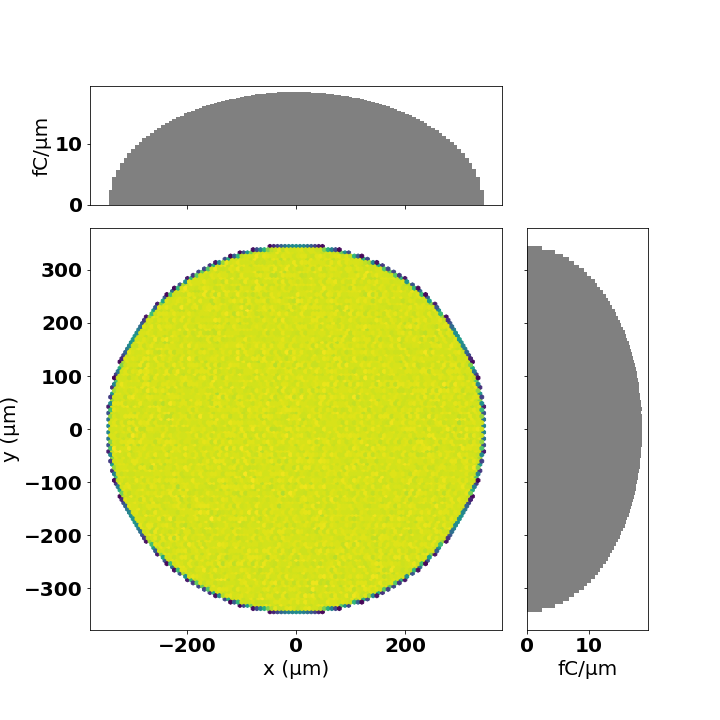}\\
 {(c) Uniform, fixed radius initial electron distribution.}&
 {(d) SG generated laser distribution.}\\
 \includegraphics[width=65mm, trim= 0 0 70 75, clip ]{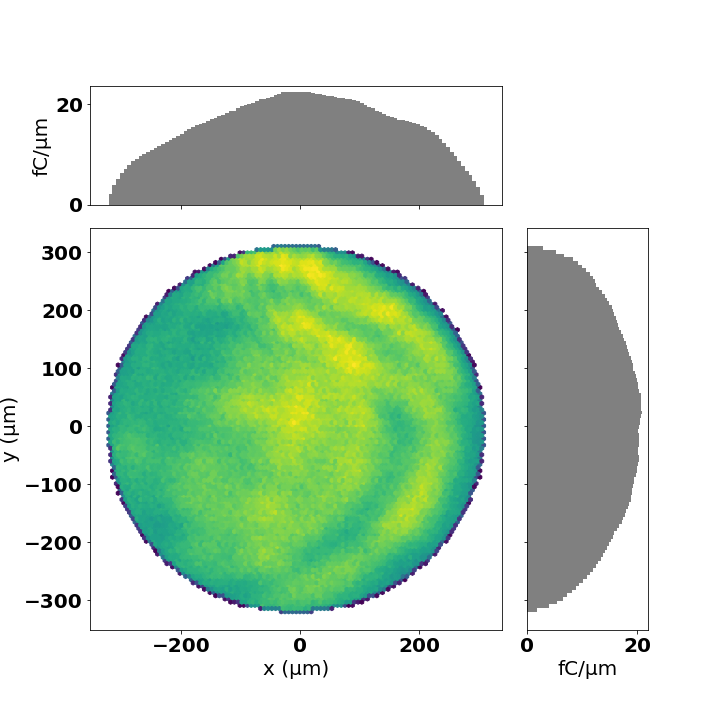} &\\
 {(e) Circular-trimmed VCC-generated initial electron distribution.} 
\end{tabular}
\caption{Shown are the various initial electron distributions that were generated to determine how sensitive Astra calculations are to the initial particle distributions. The distributions shown here represent one million macro-particles, such that the features are clearly visible. A simple convergence study confirmed there was sufficient agreement in the bulk parameters for less particles (10,000 macro-particles) while reducing simulation run time. }
\label{fig:sensitivity-laserprofs}
\end{figure*}

\subsubsection{Sensitivity of LUME-Astra Predictions to Different Laser Distributions}
We assessed the sensitivity of the simulation results on a realistic vs. idealized laser profile in simulation. In this case, sensitivity is evaluated by whether the bulk properties of importance, normalized transverse emittance and transverse beam size, differ more than 10\% from the values calculated from the uniform initial distribution. This threshold is close to the resolution of such measurements in the physical machine.

\begin{figure*}[ht!]
\begin{tabular}{cc}
  \includegraphics[width=0.5\textwidth]{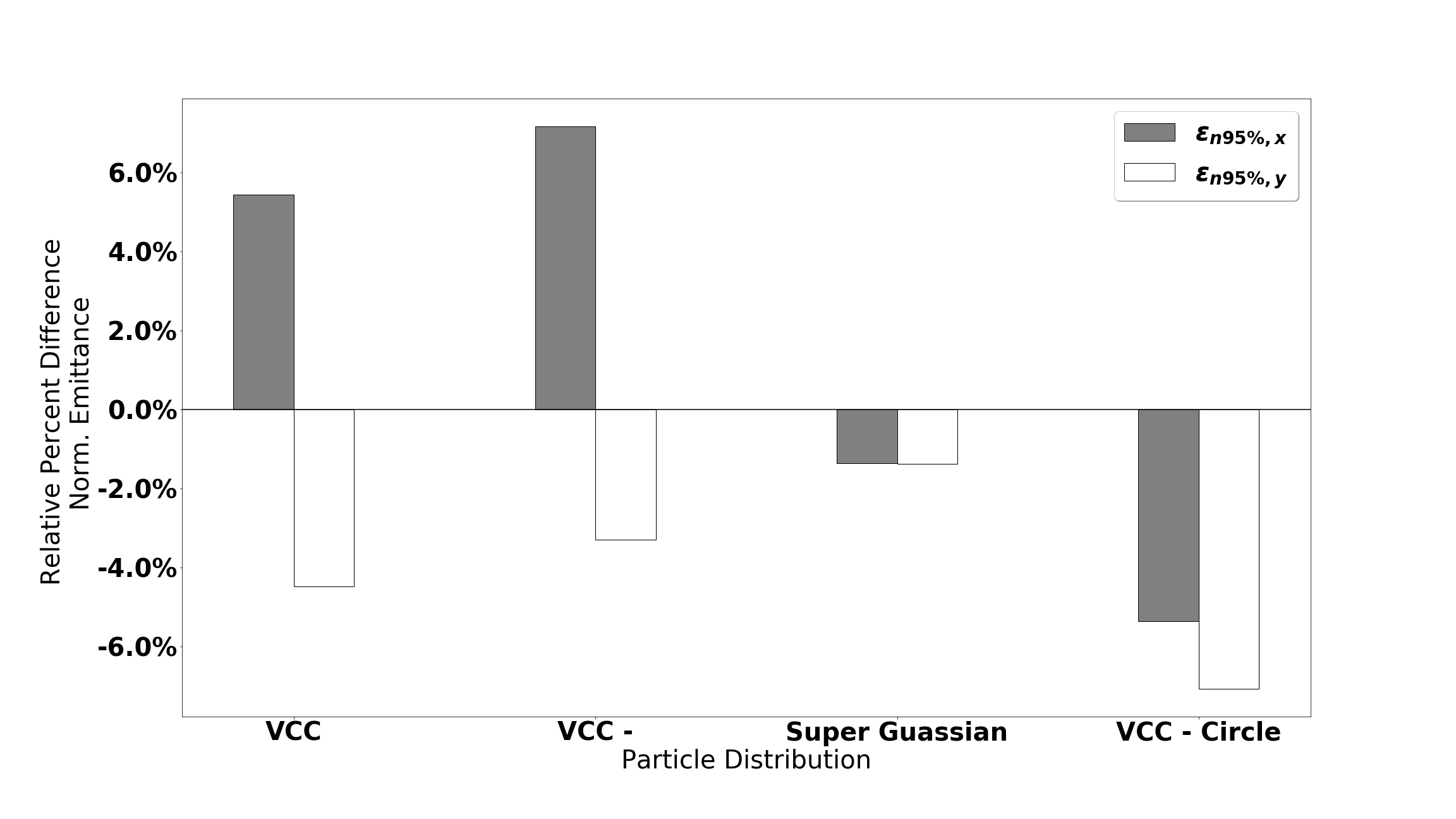} &   
  \includegraphics[width=0.5\textwidth]{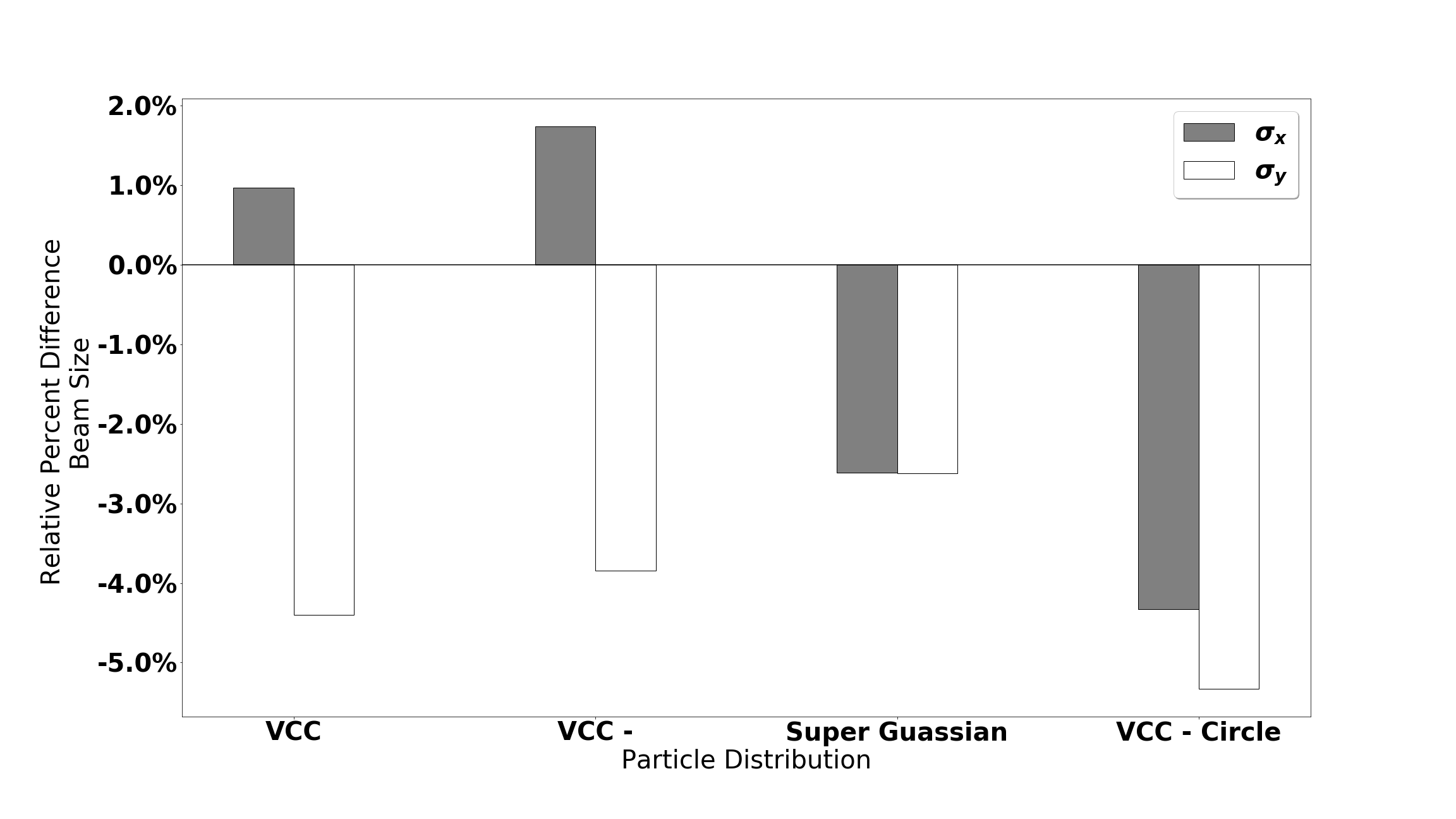} \\
(a) & (b)\\[6pt]
 \includegraphics[width=0.5\textwidth]{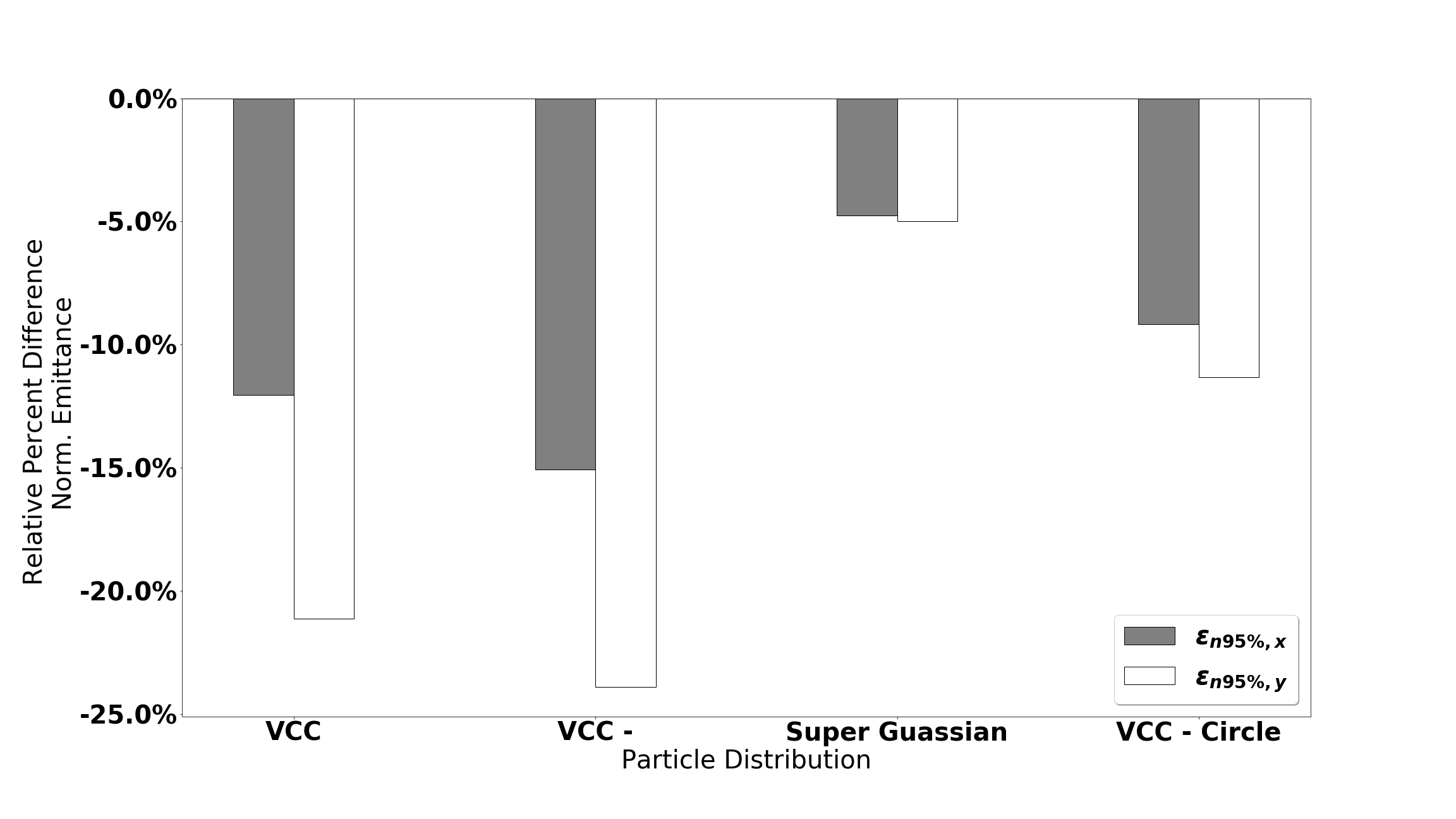} &   
 \includegraphics[width=0.5\textwidth]{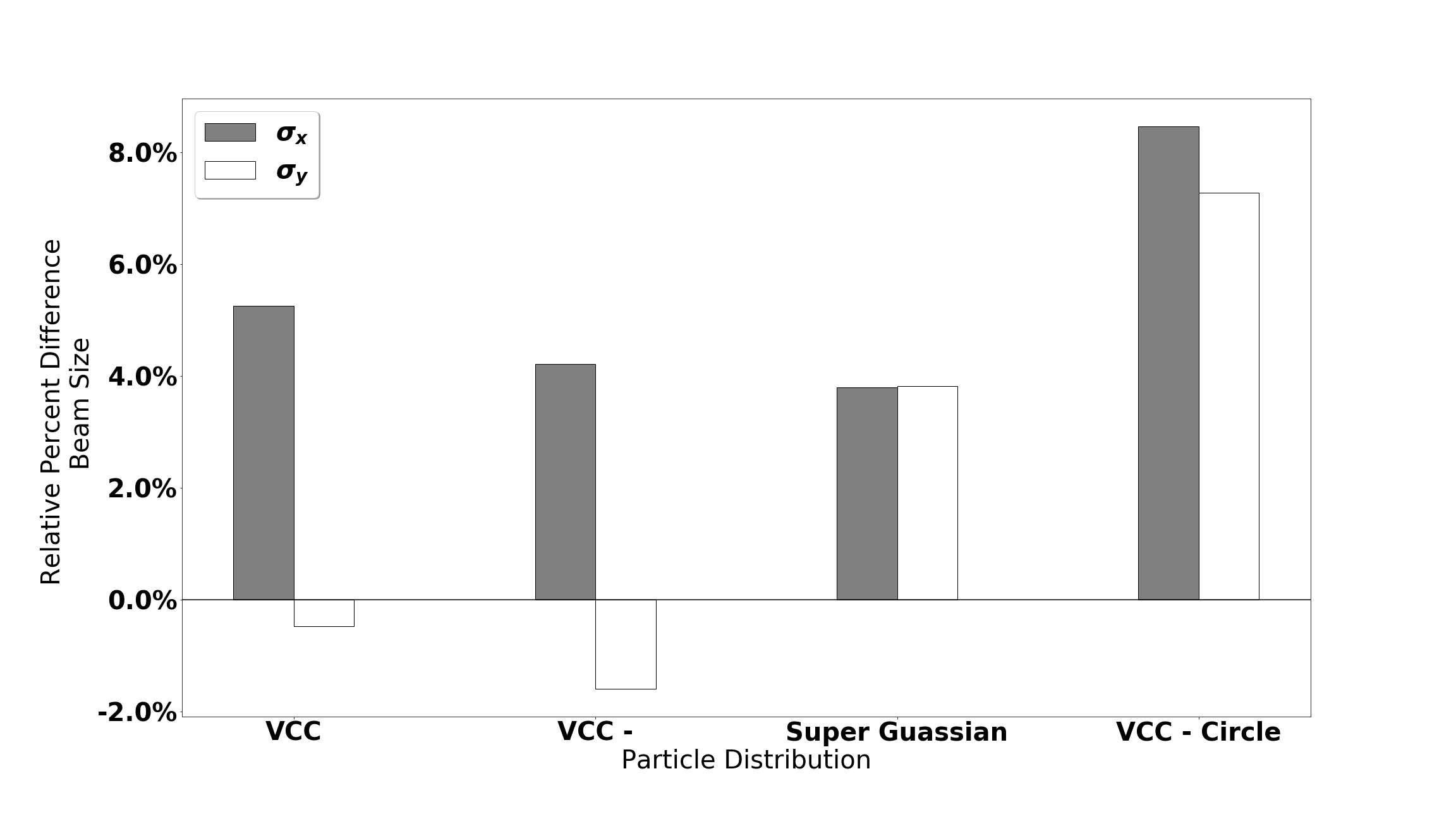} \\
(c) & (d)\\[6pt]
\end{tabular}
\caption{Shown are comparisons in the values of the end beam emittance and beam sizes, simulated by LUME-Astra for each of the laser profiles shown in Fig.~\ref{fig:sensitivity-laserprofs}, for two different bunch charges (top, 5~pc, bottom 50~pC). The percent difference is relative to the emittance or beam size as simulated from the uniform distributions (Fig.~\ref{fig:sensitivity-laserprofs}, d). This shows how much the initial particle distribution can affect the output electron parameters, and that Astra is sensitive to these different initial particle distributions. This further motivates the need to include VCC images as input to the surrogate model.}
\label{fig:sensitivity-Astra}
\end{figure*}

First, a candidate laser profile with features including rough edges, as well as fluctuations within the bulk of the laser spot, was chosen. The transverse profiles compared include: uniform radial distribution, SG distribution, and the candidate laser profile with a Gaussian blur applied. Shown in Fig.~\ref{fig:sensitivity-laserprofs} are each laser profile, with nominal charge of 10~pC total, for two different amounts of sampled particles: 1 million and 10k. The candidate VCC image is shown in Fig.~\ref{fig:sensitivity-laserprofs} $a$, as well as the particles generated from a blurred version of the candidate VCC in Fig.~\ref{fig:sensitivity-laserprofs} $b$. Having a similar, but slightly smoothed version of the candidate VCC will address whether the simulation is sensitive to the internal structure of the spot size, or just coarse features such as rough edges. This is further investigated by removing the edges of the candidate VCC images, and keeping internal features. This distribution and resulting particles are shown in Fig.~\ref{fig:sensitivity-laserprofs} $d$. 

Two highly uniform distributions were prepared for comparison as well. First, is a SG distribution. The second, which is often used as the standard distribution for simulations, is a uniform density distribution with only a maximum radius specified. These distributions are shown in Fig.~\ref{fig:sensitivity-laserprofs} $c$ and $e$. 
The temporal structure for all of the particle distributions generated for this sensitivity study, as well as for the surrogate model training, was a Gaussian with standard deviation of 8.5~ps. This time distribution was held constant for all simulations. 

For each laser profile, a particle bunch with 10k particles was generated and tracked through the injector lattice in Astra to calculate various resulting beam outputs. It was determined that the bulk parameters in the simulation can be recovered with sufficient fidelity and speed using 10k particles. The primary quantities of interest in this study were the resulting normalized transverse emittances (95\%, about two sigma, core emittance) and beam sizes. Astra simulations were completed for each laser profile at two charge settings (5~pC and 50~pC), with all other parameters, such as solenoid magnet gradient, held constant. The resulting transverse emittances and beam sizes at the YAG screen, 1.49 meters from the cathode, are shown in Fig.~\ref{fig:sensitivity-Astra}.

It is clear that the emittance and beam size from Astra simulations are sensitive to the realistic beam distributions, relative to a uniform beam distribution. For the SG distribution, which emulates an ideal flat-top beam distribution, the emittance in each direction is the same, however there is a difference seen in emittances calculated from a VCC generated laser profile. Clearly the asymmetry of the VCC generated laser distributions can be captured by the simulation, as shown by the difference in beam size and emittances in each transverse direction, for a given VCC laser profile. These results suggest that using realistic laser profiles could result in simulated training data which is sufficiently different from that generated from idealized conditions.

\section{Surrogate Model to Emulate Astra Simulations}

As stated, two major use cases for the LCLS-II surrogate model are: (1) to aid offline experiment planning and start-to-end optimization, and (2) to provide non-invasive predictions of the beam behavior, given measured upstream inputs. With those applications in mind, it is critical to evaluate the ability of the surrogate model to accurately emulate the behavior of the Astra simulations under optimization and in predicting the output beam distributions. For example, the model needs to be able to interpolate between different types of laser distributions (as seen on the VCC images) to be useful in online prediction under changing laser conditions. Here we evaluate the ability of a neural network surrogate model to interpolate to regions of parameter space not seen during training and to reliably provide accurate predictions when used in optimization.

\subsection{Data Generation and General Training Procedures} 

Training data was generated in two ways. For scalar model development, the standard particle generator in Astra was used by supplying laser radii as inputs. A large random sample of the laser radius, cathode cavity RF phase, beam charge and solenoid strength was then simulated to create a data set which assumed standard incoming beam parameters. This data was used for scalar model training.

\begin{figure}[ht!]
     \centering
     \includegraphics[width=85mm]{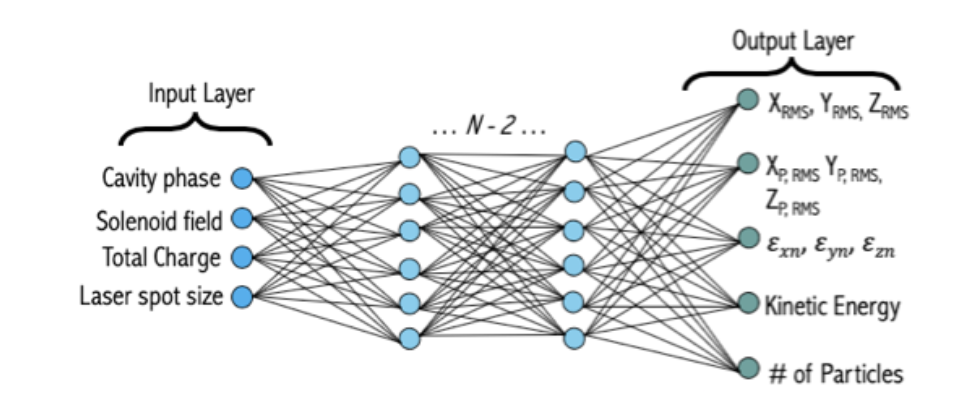}
     \caption{Feed-forward, fully-connected neural network architecture used for the scalar-to-scalar surrogate model.}
     \label{fig:FF_arch}
\end{figure}

\begin{figure}[h!]
     \centering
     \includegraphics[width=85mm]{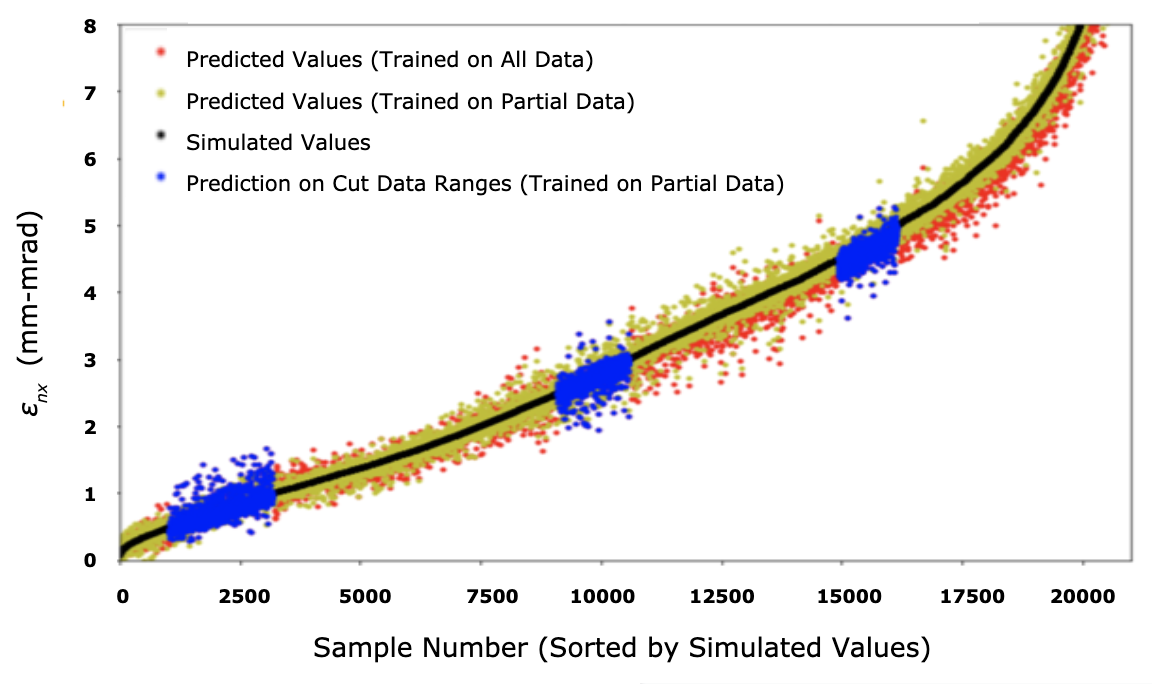}
     \caption{Performance of the surrogate model in interpolating to unseen emittance values (shown in blue).}
     \label{fig:FF_interp}
\end{figure}

\begin{figure}[h!]
     \centering
     \includegraphics[width=80mm]{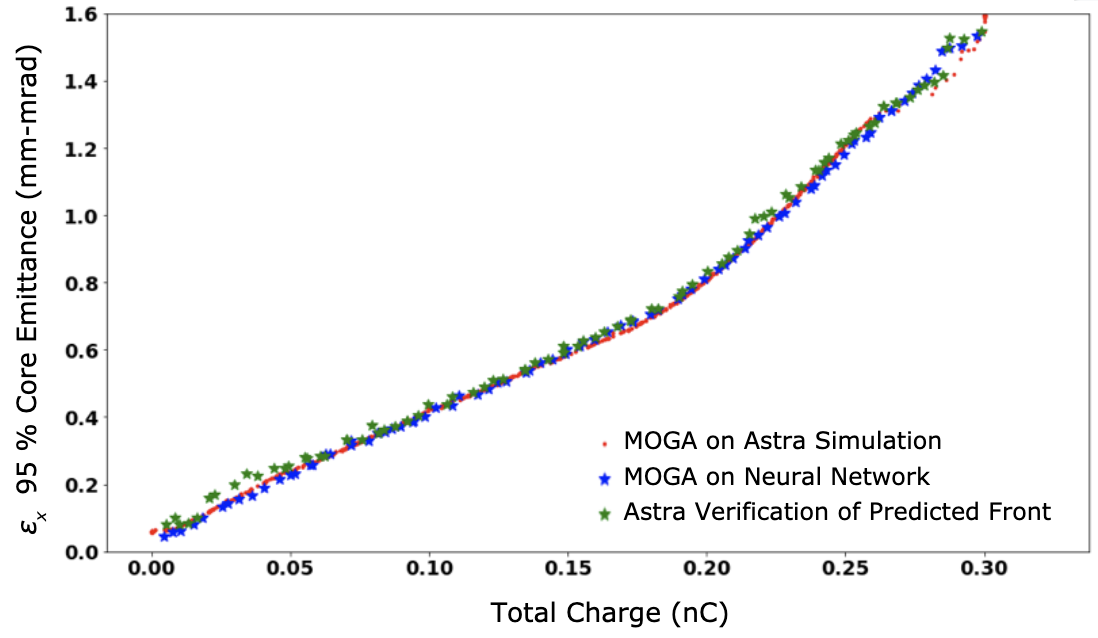}
     \caption{Result of running a standard optimization with MOGA on the surrogate model, compared with results from Astra. In this case, the objective was to maximize the beam charge and minimize the emittance. The predicted Pareto points from the surrogate model are also verified by re-running the inputs in Astra. This shows the model is reliable for use in multi-objective optimization and can be used as part of start-to-end optimizations for LCLS-II.}
     \label{fig:FF_moga}
\end{figure}

Further training data was generated by running measured VCC laser distributions and idealized SG laser distributions through LUME-Astra while randomly sampling injector input settings. For each unique laser profile, 2,000 randomly-sampled points in the input space were generated. The predicted output includes the electron beam distribution as it might be measured at a yttrium-aluminum-garnet (YAG) screen, along with bulk statistical quantities such as normalized emittance and beam sizes. Two simulated data sets consisting of approximately 60,000 samples from SG particle distributions and 70,000 from VCC measurements were generated.

\begin{figure*}[ht!]
     \centering
     \includegraphics[width=\textwidth]{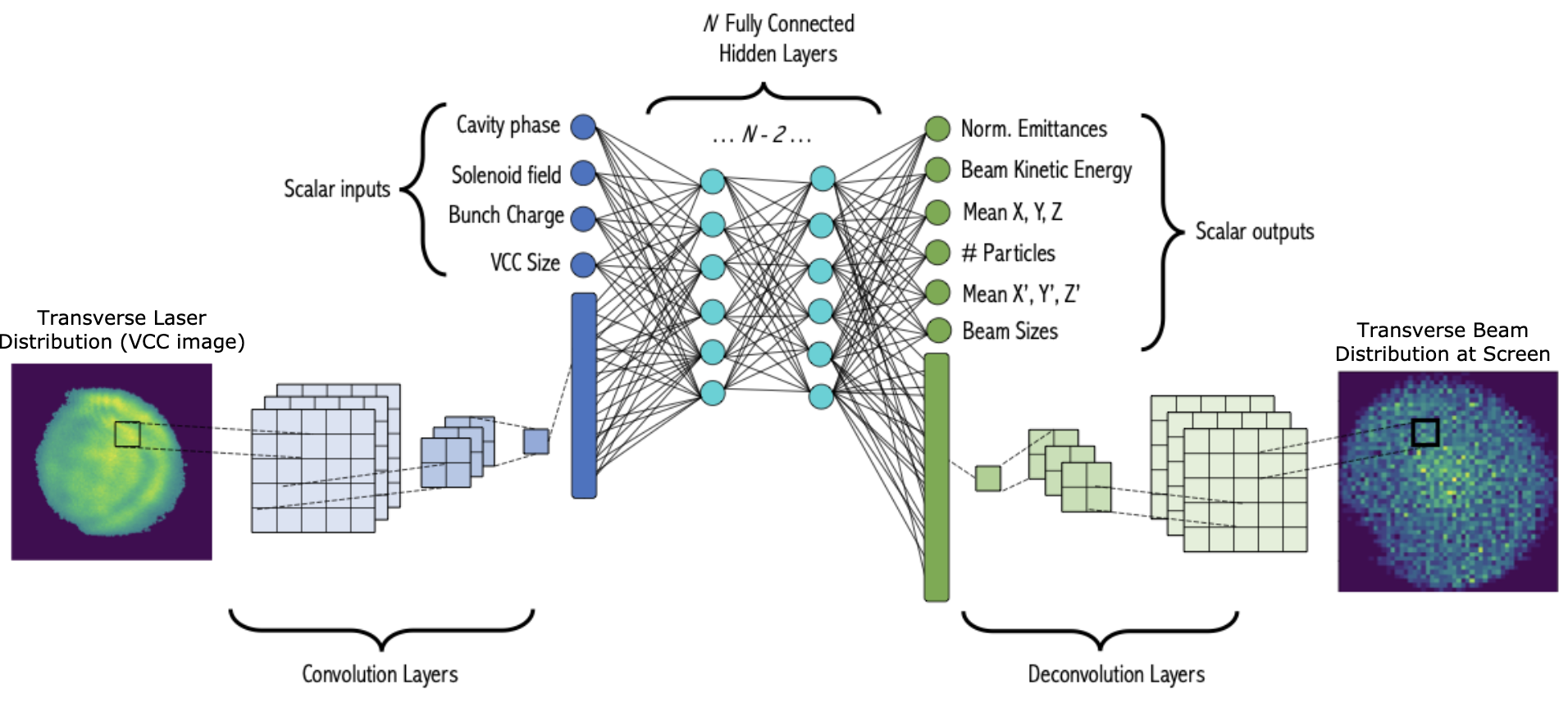}
     \caption{Encoder-Decoder CNN architecture used for prediction of beam transverse distributions and scalar beam parameters, with the VCC laser distribution as a variable input. To process the VCC images (binned into  $50 \times 50$ pixels), the encoder consists of 3 convolutional layers with 10 filters each, alternating with max pooling layers for 2 $\times$ downsampling. The scalar input settings are concatenated into the first of 4 fully-connected layers in between the encoder and decoder. The scalar outputs are obtained from the last of these layers. Finally the decoder CNN consists of 3 convolutional layers alternating with 2 $\times$ upsampling layers, resulting in an output transverse beam prediction image with $50 \times 50$  bins. }
     \label{fig:CNN_img_arch}
\end{figure*}

All neural network model development and training was done using the TensorFlow and Keras libraries \cite{chollet2015keras}. Each model was trained by minimizing a mean squared error loss function, using the Adam optimization algorithm \cite{kingma2015adam}. Several different neural network architectures are used, as described in the following sections. During any training process, the training samples are used for fitting the model. The  validation loss is calculated and monitored during training to avoid overfitting, but is not included directly in the weight updates. All testing samples are withheld from the training process entirely. 

\subsection{Scalar Model Performance in Interpolation and Multi-objective Optimization}

Here we assessed the performance of a model that uses only the laser radius of a uniform laser distribution as an input rather than a laser distribution images. Typically a laser radius with a static profile is used for most multi-objective optimization studies on injectors (including the LCLS-II injector). We predict a wide variety of output scalars that are relevant for optimization studies. Fig.~\ref{fig:FF_arch} shows the basic inputs and outputs. The neural network architecture itself consisted of 8 layers (6 hidden layers), each using a hyperbolic tangent activation function. The hidden layers each had 20 nodes; the input layer had 4 corresponding with each input and the model output 16 scalar predictions.

The performance on the scalar predictions is similar to that  for the CNN case, described later and shown in  Fig.~\ref{fig:CNN_scalar}. To ensure the model can be used for optimization studies, we first left out sections of the parameter space from both the training and the validation set to verify the model can interpolate accurately (see Fig.~\ref{fig:FF_interp}). Next, we also verified that optimization with a standard multi-objective genetic algorithm (MOGA) on the model produces an accurate Pareto front (see Fig.~\ref{fig:FF_moga}). We use a similar MOGA setup to that described in \cite{PhysRevAccelBeams.23.044601}. For the verification, we run the input settings from the predicted front in Astra. Based on these results, this scalar version of the injector surrogate model can already be used as a component in start-to-end optimization of new setups for LCLS-II (e.g. by replacing the simulation of the gun).

\subsection{Performance for Beam Distribution Predictions and Interpolation to New (Out-of-Distribution) VCC Images}

For prediction of the transverse beam distributions and scalar outputs and taking into account the VCC image, we introduced an encoder-decoder style CNN architecture (shown in Fig.~\ref{fig:CNN_img_arch}); this approach has not been taken  before for injector surrogate modeling. Each output transverse distribution is binned into a $50 \times 50$ image. Sixteen scalar beam parameter outputs and the scalar extents of the beam distributions are predicted as well. For each VCC image, a random sample of the input settings was conducted in Astra. In this case, the model is trained on 60,320 samples, with 7,540 samples held out for training and testing.

 \begin{figure*}[ht!]
     \centering
     \includegraphics[width=180mm]{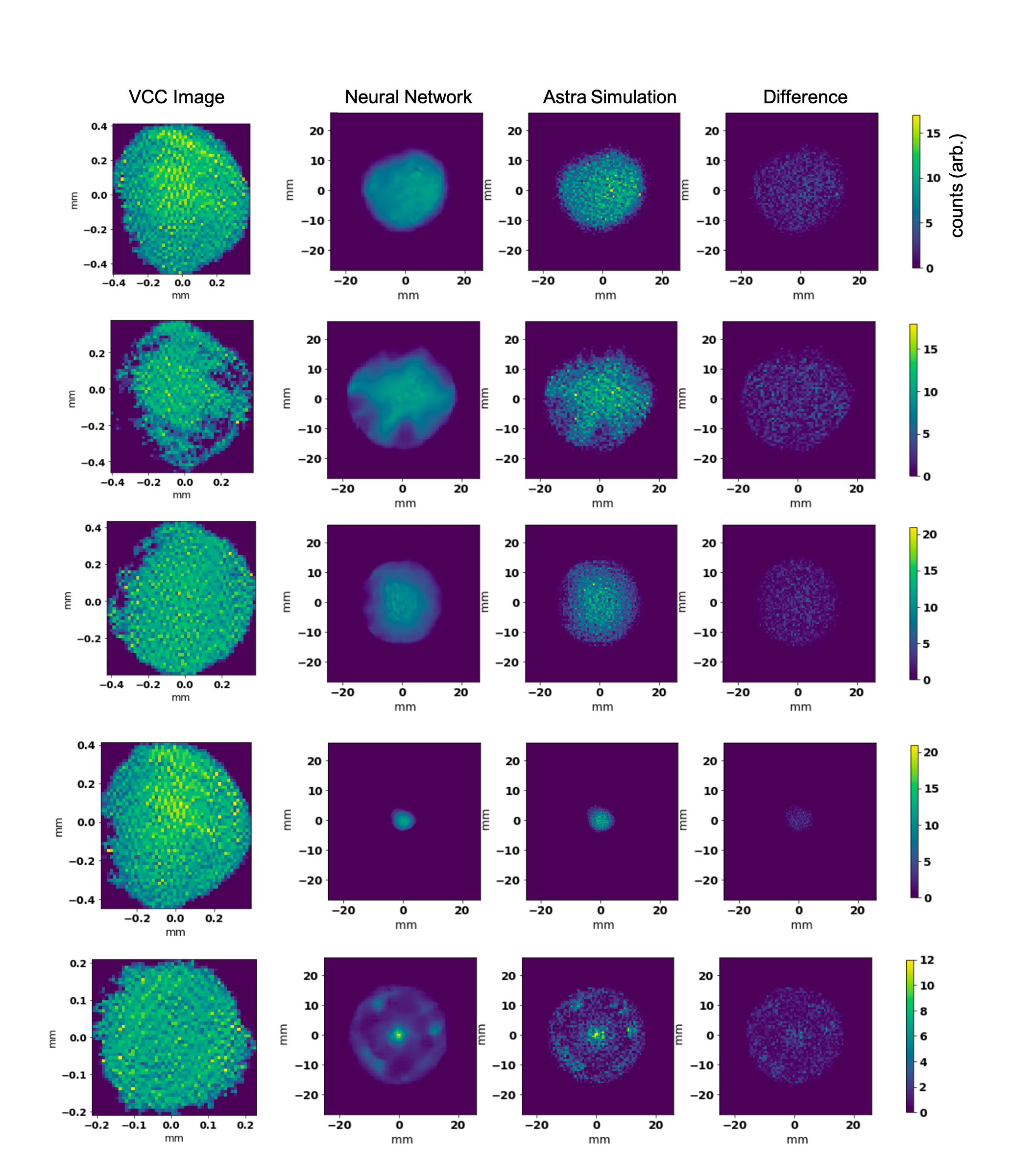}
     \caption{Examples of the neural network predictions and Astra simulation results for the transverse beam distributions. The corresponding VCC inputs used in each case are shown at left. The agreement is good, even for cases with irregular beam distributions. This demonstrates that the model can interpolate between measured input laser distributions (as seen on the VCC) and provide realistic predictions of the expected transverse beam distribution from simulation. This is important for using this model online in the accelerator, as the initial beam distribution will vary with time.
     }
     \label{fig:Img_pred_single}
 \end{figure*}
 
    \begin{figure}[ht!]
     \centering
     \includegraphics[width=90mm]{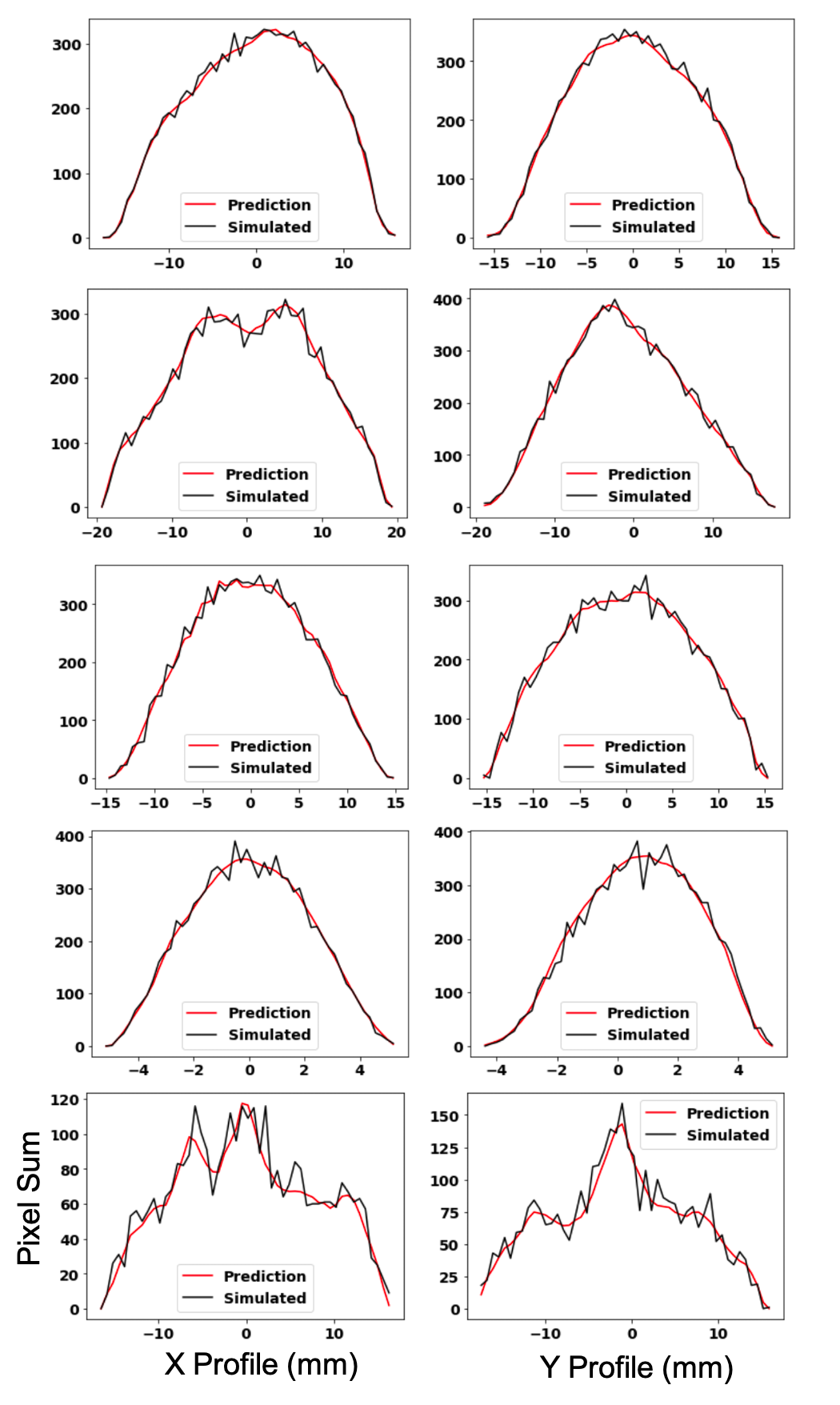}
     
     \caption{Predicted and simulated profiles, for the same cases shown in Fig.~\ref{fig:Img_pred_single}.}
     \label{fig:Img_pred_profile}
 \end{figure}

\begin{figure}[ht!]
     \includegraphics[width=60mm]{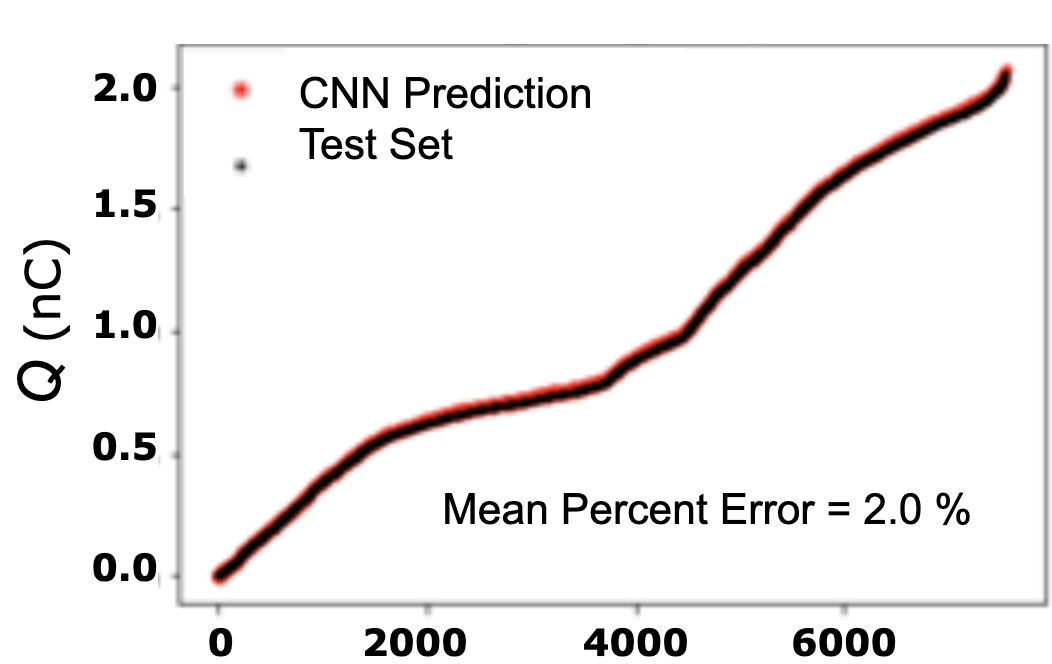}
     \includegraphics[width=60mm]{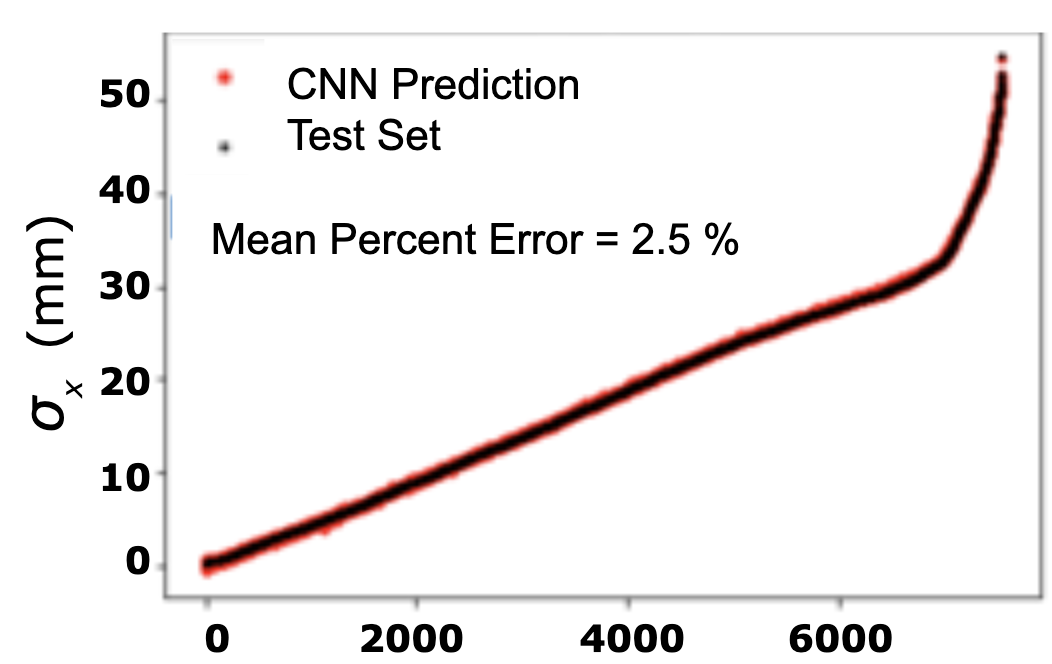}
     \includegraphics[width=60mm]{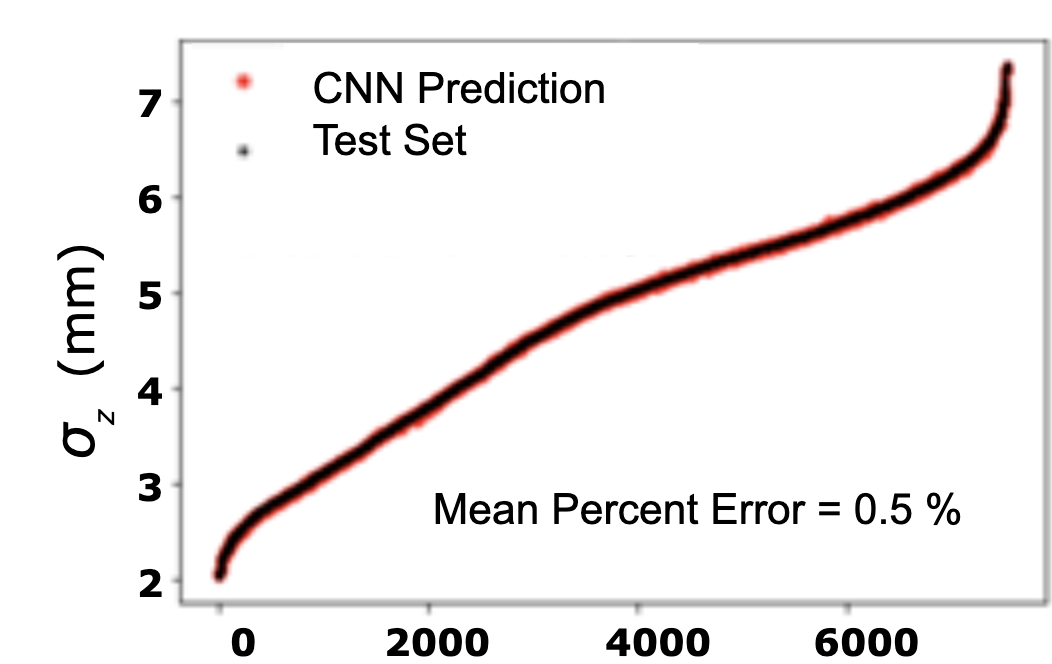}
     \includegraphics[width=60mm]{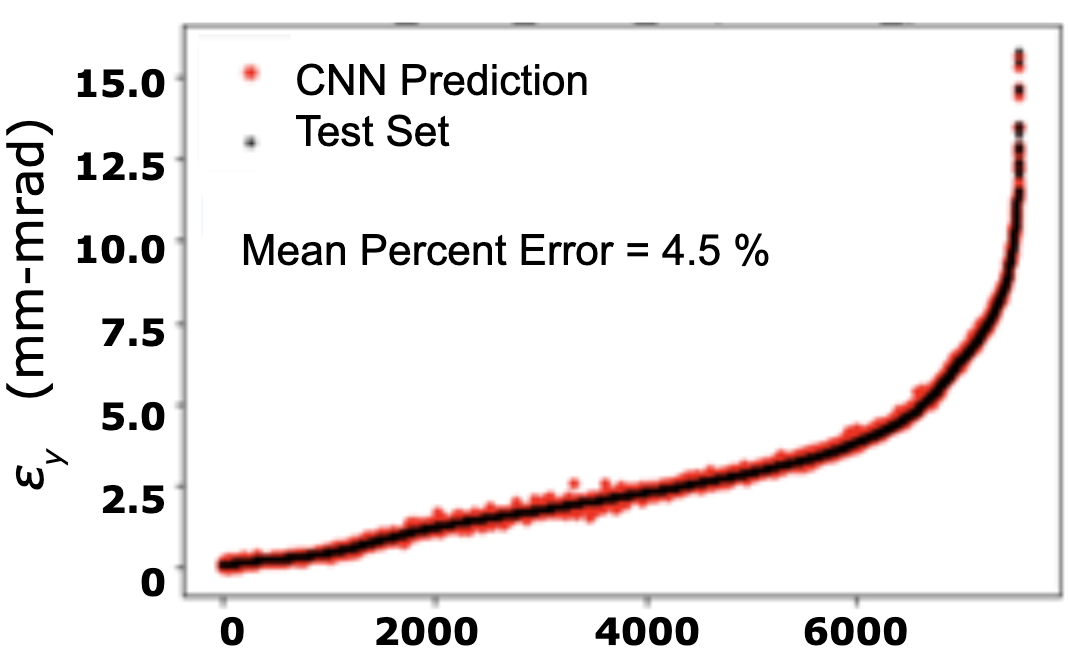}
     \includegraphics[width=60mm]{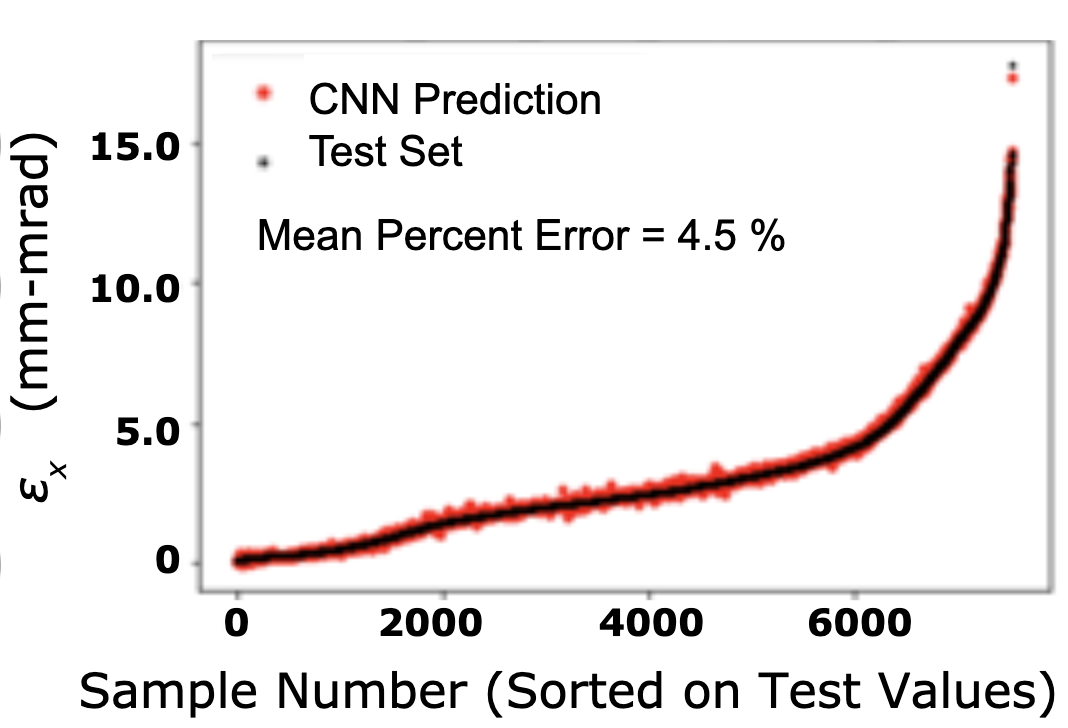}
     \caption{Example of prediction performance of surrogate model on scalar output parameters of interest.}
    \label{fig:CNN_scalar}
\end{figure}

To assess the ability of the model to interpolate between different laser distributions (so that it can provide accurate predictions on new VCC images as the laser distribution shifts over time), we selected a set of VCC images that had patches of intensity within the bulk of the laser spot missing, and held this data out of the training and validation set. We find good agreement between simulation results and the surrogate model predictions, even for cases with irregular beam distributions (see Fig.~\ref{fig:Img_pred_single} and ~\ref{fig:Img_pred_profile}). This indicates that the model can be used online with the running accelerator to provide non-invasive estimates of the transverse beam profile (i.e. as both an online model of the injector and a virtual diagnostic), similar to how an online physics simulator could, but with much faster-to-execute predictions. The performance of the model on bulk scalar predictions is shown in Fig.~\ref{fig:CNN_scalar}.
 
\section{Transfer Learning}

The previous sections demonstrated the ability of the surrogate model to reliably emulate predictions from Astra simulations. However, the issue of how these predictions compare to measured beam parameters remain. Because we have very little measured data, we generated an initial model trained on simulation data and then modify it to be consistent with measured data afterward. Here, we develop and demonstrate a transfer learning procedure to accomplish this.

Transfer learning encompasses a broad class of machine learning approaches wherein the performance of a model at a particular task or domain may be improved by transferring information from another related but different task or domain \cite{pan2009survey}.  In traditional approaches to machine learning, the distribution over feature space and the distribution in target space must be identical during training and deployment. If any such differences, termed distribution shifts, exist, the performance of the trained model is severely degraded \cite{shimodaira2000improving}. Transfer learning is thus one approach to handling distribution shifts between target domains (e.g. simulation to measurements, idealized laser beam shapes to non-idealized ones), and it has been successfully applied to diverse applications including image classification \cite{kulis2011you}, anomaly detection \cite{nam2017heterogeneous}, text sentiment analysis \cite{wang2011heterogeneous}, etc.

\begin{figure*}[ht!]
    \centering
    \includegraphics[width = 0.9\textwidth]{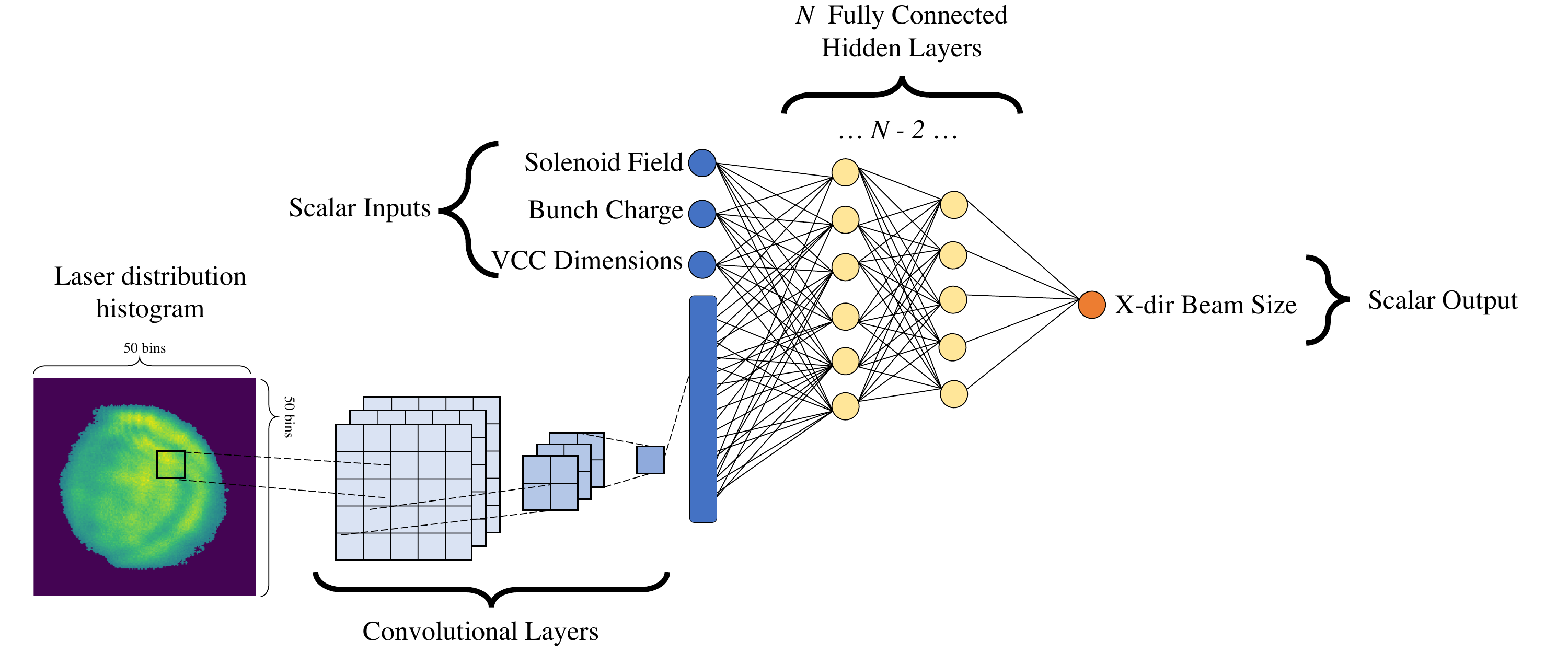}
    \caption{This schematic shows the surrogate model architecture used for transfer learning between the simulation and measurement domain. Scalar settings and a histogram of the laser distributions are used to predict scalar output values including emittances and beam sizes.}
    \label{fig:schematic}
\end{figure*}

To find a suitable transfer learning approach for this class of accelerator surrogate model, we first prototyped the approach on simulation data. We started with a primary model trained on SG beam distributions and then expand the model training to include the simulation results obtained from measured VCC images. We then applied the same procedure to the measured data.

A base model was trained on simulation data set until the mean-squared error (MSE) loss did not decrease over several training epochs. Then all of the model weights except the weights and bias values for the last two layers were frozen. A dropout layer \cite{JMLR:v15:srivastava14a} was also included between the frozen and unfrozen layers to reduce over-fitting. The initial learning rate was then set at \SI{5e-4}, and decreased on an exponential schedule. The training was terminated using an early stopping method. After this, the learning rate was reduced by two orders of magnitude, and all of the weights and bias values in the model were trained, referred to as annealing. Models at each of these stages, after the initial transfer learning (referred to as TL models) and after annealing (referred to as TL + A models) are used to compare predictions and the resulting loss accuracy of the prediction.

\subsection{Data Creation and Neural Network Architecture}

The training data was generated by using the measured VCC laser distributions as inputs and generating Astra simulation output by randomly sampling injector input settings. Thus, for each unique SG laser profile and separately the VCC profiles, 2,000 randomly-sampled points in the input space were generated and the corresponding electron outputs were simulated using LUME-Astra. Each simulation calculated not only the electron beam distribution as it might be measured at a YAG screen, but also bulk statistical quantities such as normalized emittance and beam sizes. Thus, a data set consisting of approximately 60,000 samples from SG particle distributions, and 70,000 from VCC measurements were generated.

Ample simulation data was generated, and a sparse sub-sample was used for surrogate model training, which can ensure the model is able to interpolate well, and minimize over-fitting. Sub-sampling was also used to emulate small amounts of data for re-training (as one would typically have for measured data sets on an accelerator). 

The model architecture described, and shown in Fig.~\ref{fig:schematic}, depicts the general architecture of the NN surrogate models. All models took scalar settings for the solenoid value and charge as inputs, along with the 2-dimensional histogram representation of the laser intensity on the cathode. The laser distributions were $50 \times 50$ bins. The size of the laser distribution were given as the horizontal and vertical extents of the histogram, relative to the center of the histogram. These six scalar values and the $50 \times 50$ bin laser distribution are considered inputs to the models. The binned images were input into convolutional layers. Three convolutional layers with 10 $4 \times 4$ filters each are applied to the image inputs. The resulting nodes are then fed to  densely-connected layers. The densely-connected part of the network consists of 6 hidden layers with 1024, 512, 256, 64, 32, 16, 6 neurons respectively. The output layer consists of one node predicting the transverse beam size in x.

As before, all neural network model development and training was done using the TensorFlow and Keras libraries \cite{chollet2015keras}. Each model was trained by minimizing a mean squared error loss (MSE) function, using the Adam optimization algorithm \cite{kingma2015adam}.

\begin{figure}[ht!]
    \centering
    \includegraphics[width=80mm, trim= 65 90 70 90, clip ]{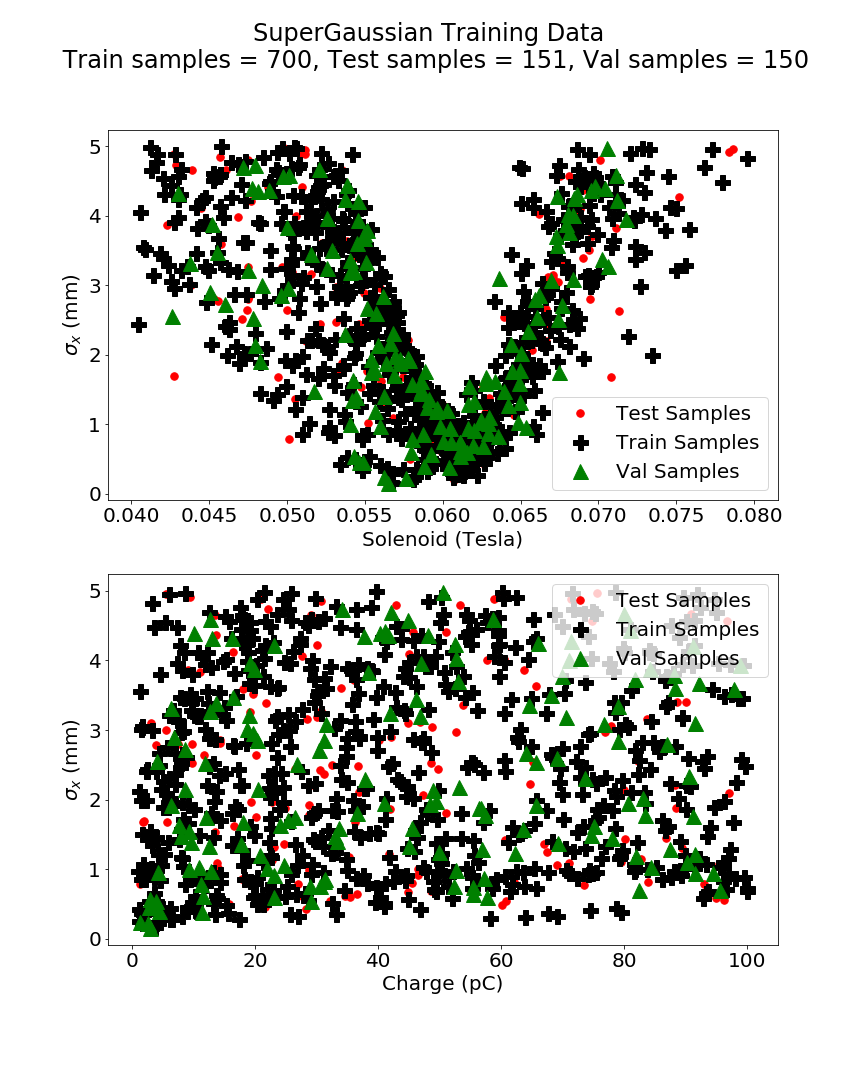}
    \caption{Shown are distribution of training, validation, and testing samples used to train the base simulation model. The data cover the scalar input parameter range, with a variety of SG laser distributions. There were 700 training samples, 150 validation samples, and 151 test samples (this is down-sampled data, ensuring the parameter space is not over-sampled). }
    \label{fig:SG-traindata}
\end{figure}

\begin{figure}[h!]
    \centering
    \includegraphics[width=80mm, trim= 60 0 30 75, clip ]{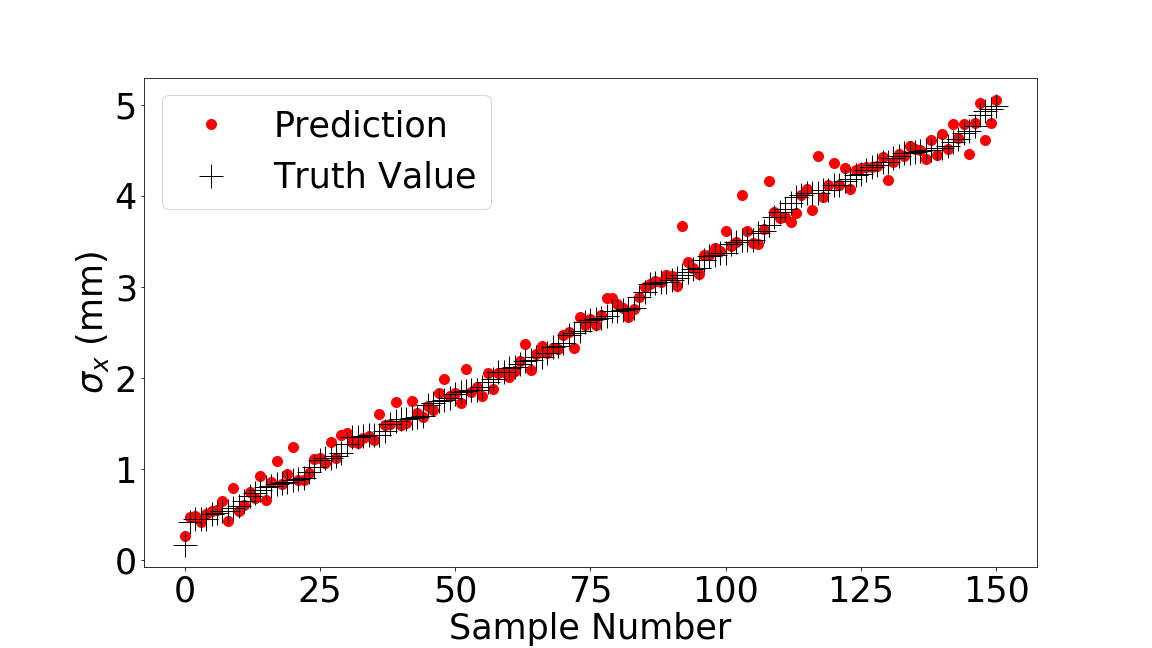}
    \caption{Predictions of the base model on test samples, which were withheld from training, sorted on magnitude. The MAPE for the test samples is 5.21\%.}
    \label{fig:SG-Predictions}
\end{figure}

\begin{figure}[h!]
    \centering
    \includegraphics[width=80mm]{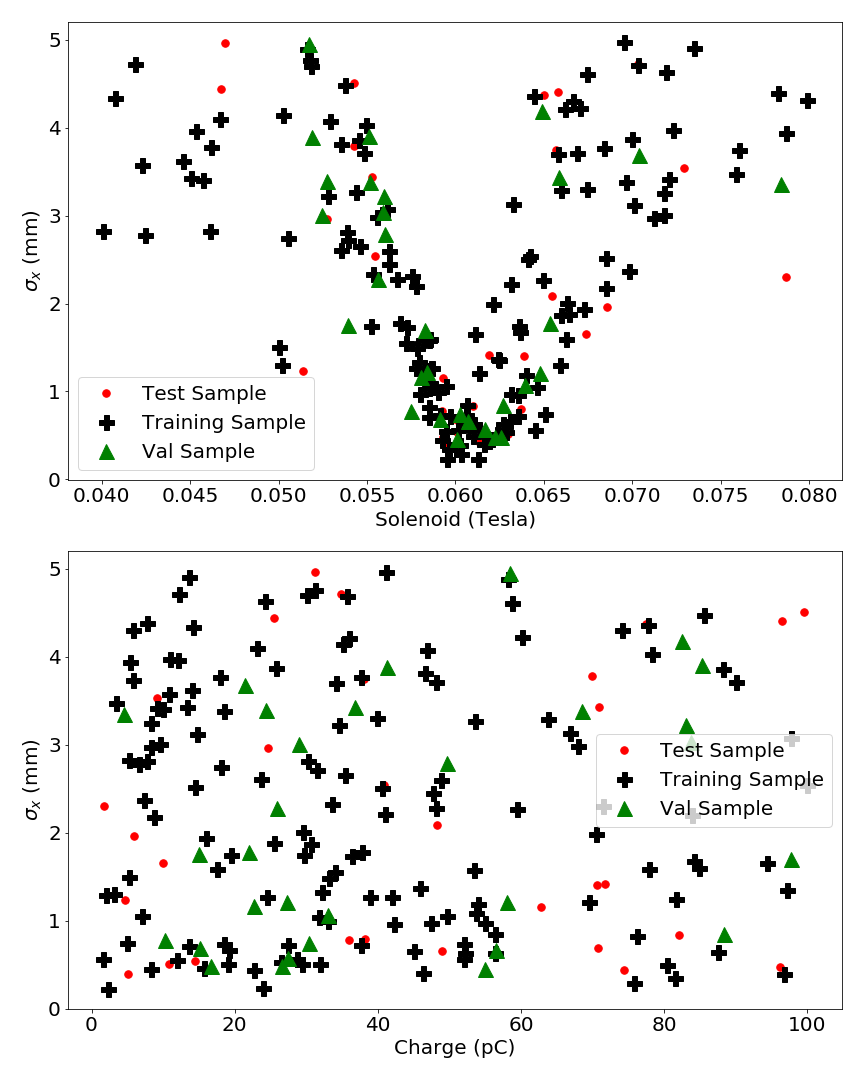}
    \caption{Shown are distribution of training, validation, and testing samples used to emulate a small, measured data set. These are simulation samples are generated with measured VCC laser distributions. There were 140 training samples, and 30 validation and test samples respectively.}
    \label{fig:h-traindata}
\end{figure}

\subsection{Transfer Learning in Simulation: Idealized Laser Distributions to Measured Laser Distributions}

We prototyped the transfer learning approach by training a model on SG laser distributions, then retraining the neural network model to predict VCC-based simulation results. Since this is a major potential source of disagreement between idealized simulations and the as-built injector, it enabled us to refine the approach prior to applying it to the measured data.

The SG generated data was used as the primary training data set, shown in Fig.~\ref{fig:SG-traindata}. The data set was down-sampled to 700 training, 150 validation, and 151 test samples, which provided sufficiently sparse coverage to ensure we were not oversampling the parameter space. The resulting predictions are shown in Fig.~\ref{fig:SG-Predictions}. To evaluate the accuracy of the models, the mean absolute percent error (MAPE) was calculated as shown: 
\begin{equation}
    \textrm{MAPE} = \textrm{mean}(\frac{|y_{true}-y_{pred}|}{y_{true}})
\end{equation}

The final MAPE on the test values was 5.21\%. Then, the VCC-based data set was down-sampled randomly to represent the small amount of measured data available. The down-sampled data is shown in Fig.~\ref{fig:h-traindata}.

\begin{figure}[htbp]
    \centering
    \includegraphics[width=80mm]{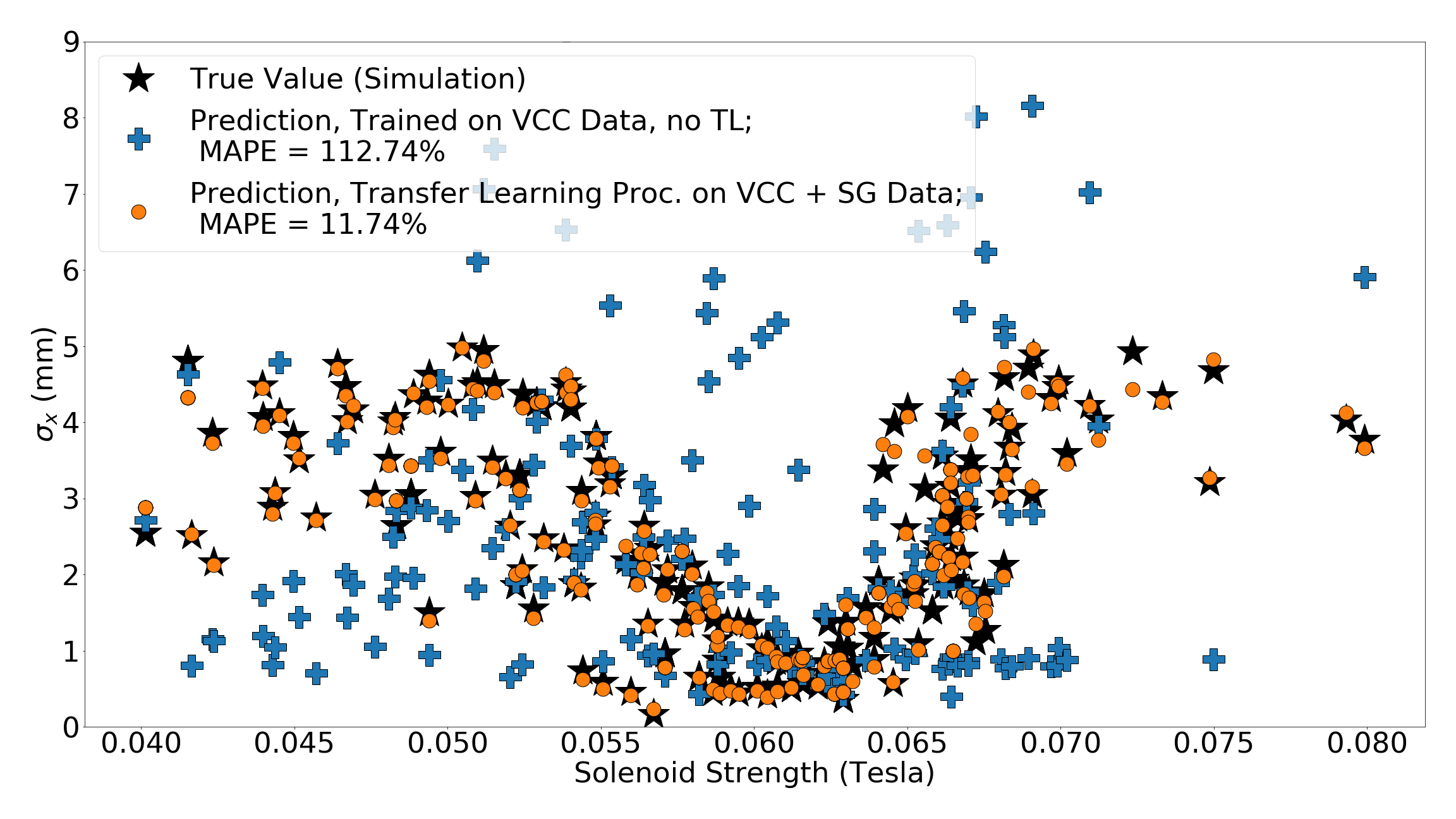}
    \caption{Transfer learning result in simulation, adapting from idealized laser distributions to measured distributions. Predictions of beam sizes are shown from a model trained only on measured VCC laser profiles without transfer learning, and from a model after transfer learning from idealized to measured profiles. The true values from simulation are sorted by the solenoid input value and represent the combined (idealized and measured) data.The performance of the model after transfer learning has better accuracy than a model trained soley on VCC-based data. This means the TL model can provide accurate predictions for a broader range of input parameters.}
    \label{fig:sgh-solscan}
\end{figure}

\begin{figure}[htbp]
    \centering
    \includegraphics[width=80mm]{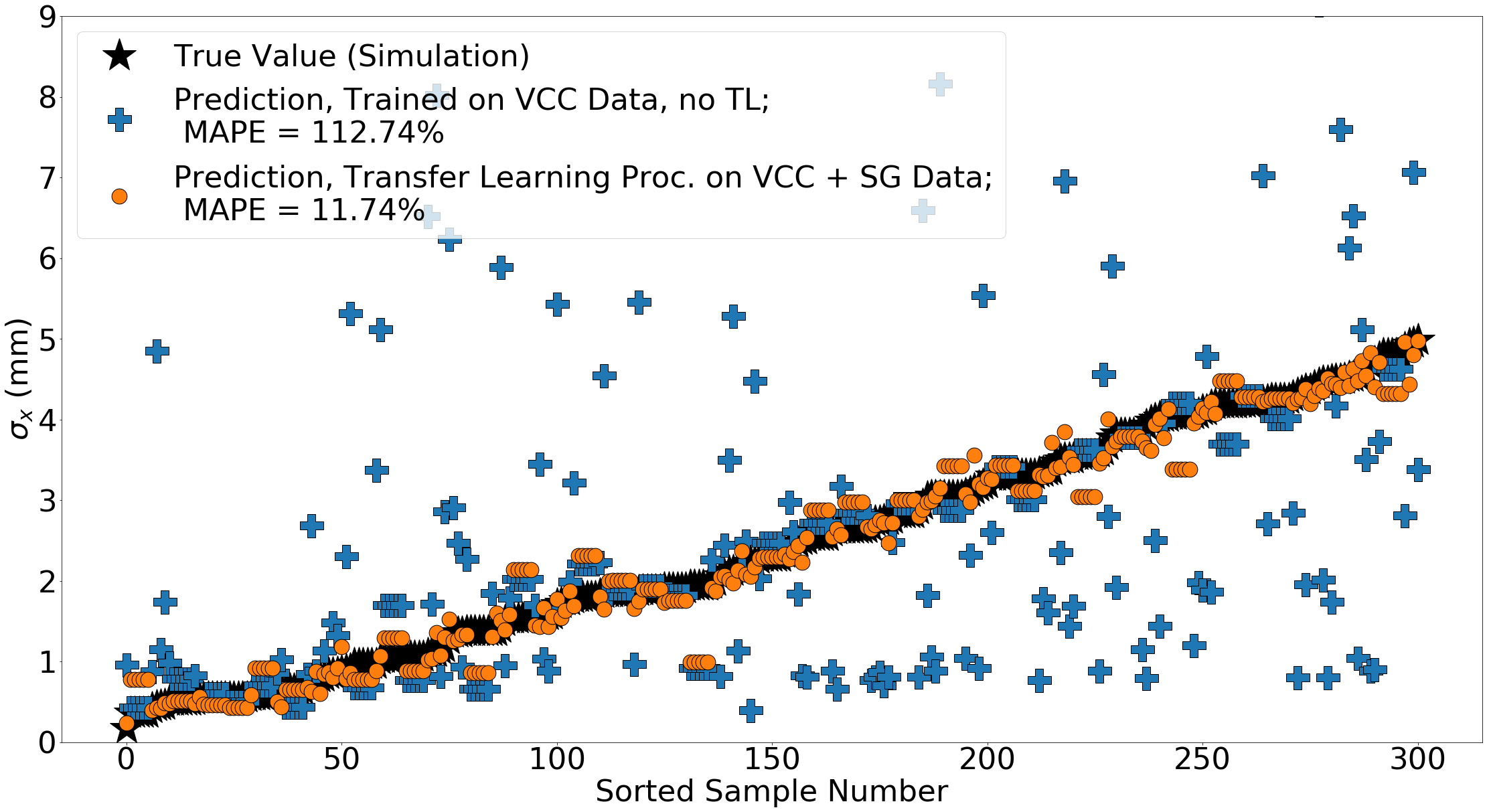}
    \caption{Transfer learning result in simulation, adapting from idealized laser distributions to measured distributions. Predictions of beam sizes from a model trained only on VCC laser profiles without transfer learning and a model after transfer learning with the combined data set was applied. The true values from simulation are sorted by the beam size magnitude.  }
    \label{fig:sgh-sorted}
\end{figure}

In this case, after training the base model, we combine the training data sets such that the neural network is trained on both simulation data sets simultaneously (but, as described earlier, with more limited adaption of the model allowed). Because the training data for the base model is 5 times larger, the smaller data set was repeated 5 times in order to create proportionally equal representation in the data set (a standard practice when dealing with imbalanced data sets). The performance on the combined test set (the test samples from the SG data set, and the VCC-generated data) is shown in Fig.~\ref{fig:sgh-solscan} and Fig.~\ref{fig:sgh-sorted}. In this case, we see that the model trained only on VCC data cannot predict the combined distribution as well as the model which underwent the transfer learning procedure.


\subsection{Transfer Learning to Measured Data}

\begin{figure}[htbp]
    \centering
    \includegraphics[width=80mm]{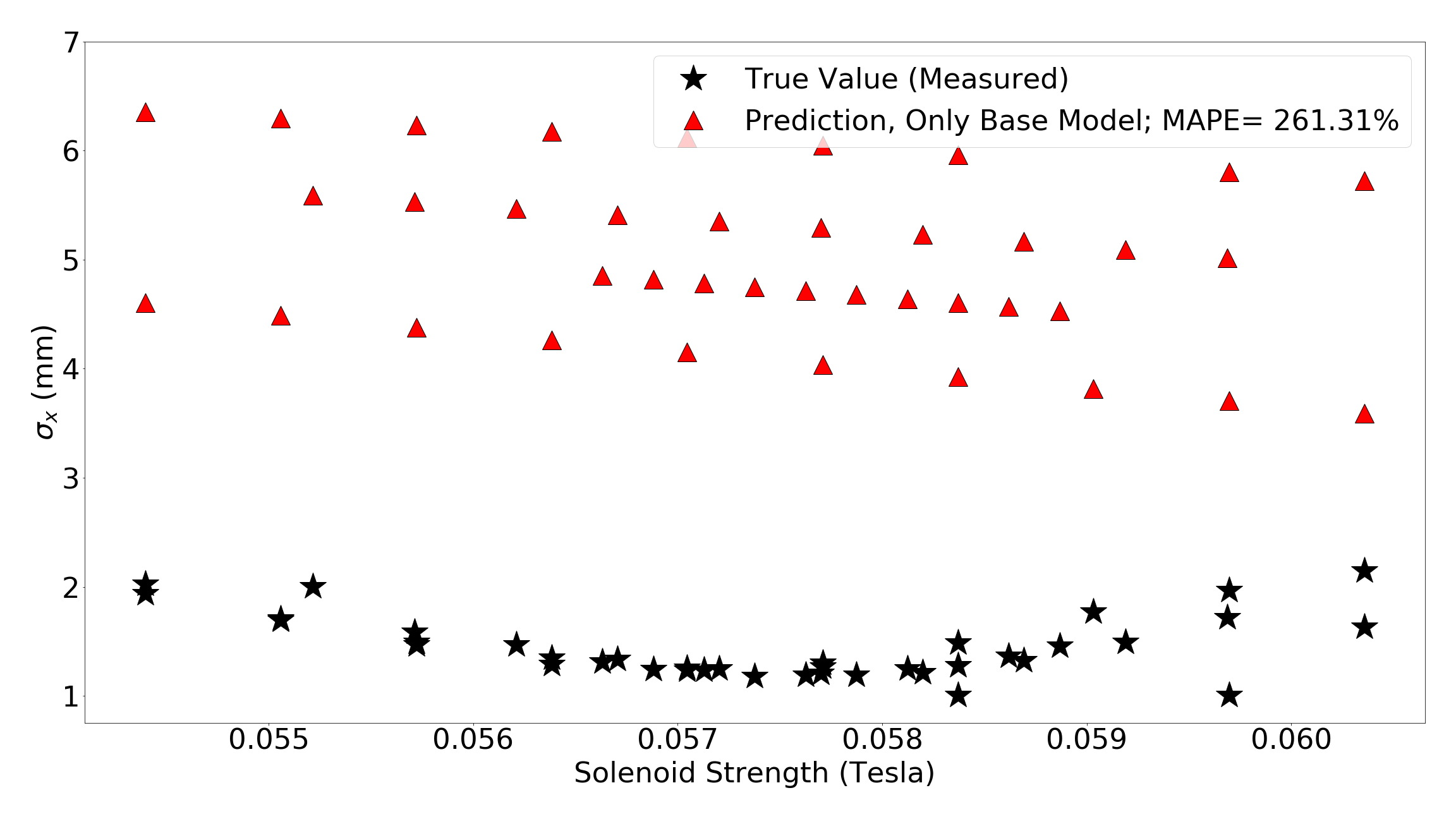}
    \caption{Predictions of measured data beam sizes from the simulation model (updated with measured VCC images). Despite the excellent performance on the simulated data, it is clear that the model trained only on simulation does not predict the measured beam sizes well.}
    \label{fig:base-only}
\end{figure}

It is clear that a transfer learning procedure is necessary. The performance of the best model trained on Super Gaussian and VCC-generated data (such as the model shown in Fig.~\ref{fig:sgh-solscan}) does not predict the measured data sufficiently well,  as shown in fig. \ref{fig:base-only}.

\begin{figure}[t]
    \centering
    \includegraphics[width=80mm,  trim = 70 70 70 70 clip]{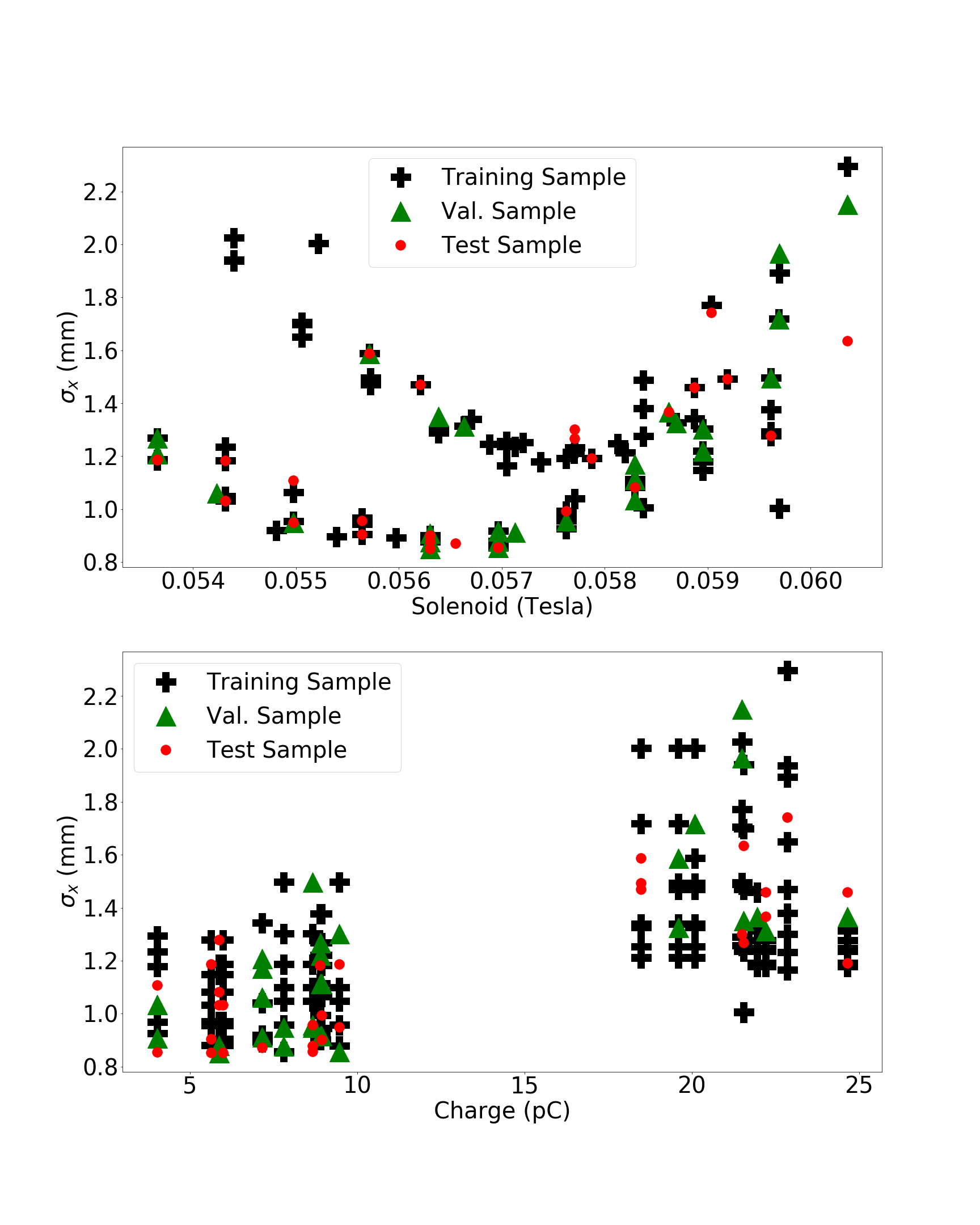}
    \caption{Training, validation, and testing samples for a surrogate model trained with only measured data are shown. There were 138 training samples, 30 validation samples, and 30 testing samples.}
    \label{fig:monly-trainingdata}
\end{figure}

\begin{figure}[t]
    \centering
    \includegraphics[width=80mm, trim = 60 30 50 30 clip]{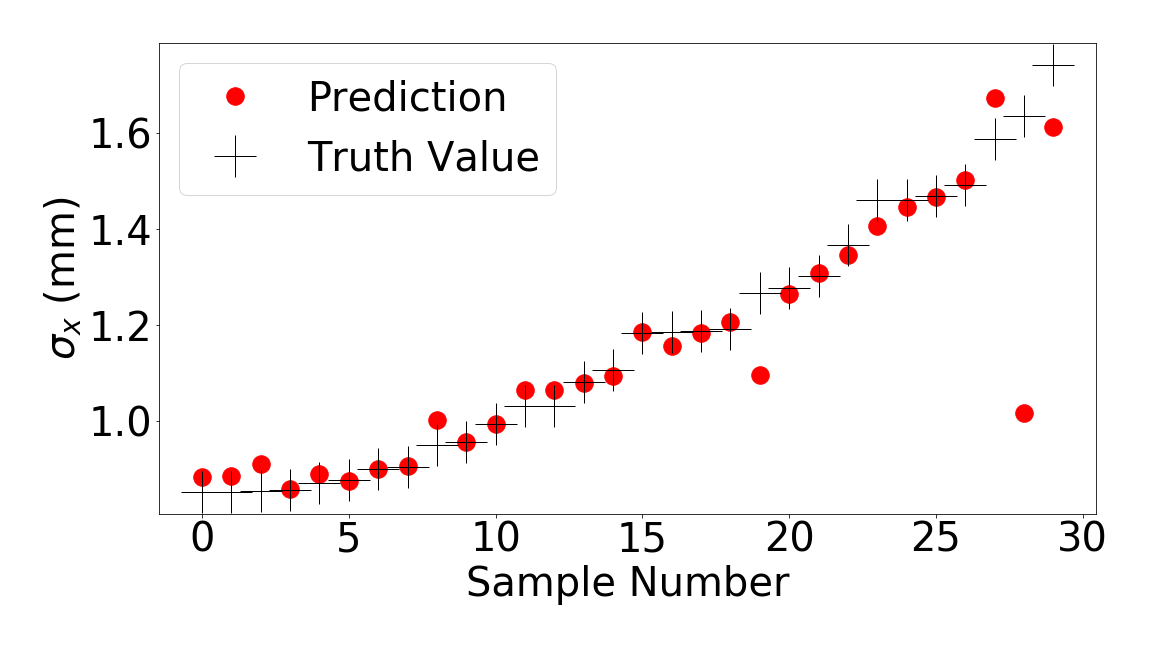}
    \caption{Predictions from a model trained on measurement data, with MAPE of 3.58\%. The test samples are shown in Fig.~\ref{fig:monly-trainingdata}.}
    \label{fig:monly-prediction}
\end{figure}

First, using the measured data collected, a simple surrogate model was trained for comparison with the transfer learning case. The training, validation and testing samples are shown in Fig.~\ref{fig:monly-trainingdata}, where the test samples are selected from the same distribution as the training and validation data. The neural network model was able to learn from the limited data and successfully predict test samples with MAPE of 3.6\%. The results are shown in Fig.~\ref{fig:monly-prediction}. However, this measured data does not present the full range of operational values. Thus, if this model was used to predict beam size at a higher bunch charge, it would likely perform poorly relative to the true machine output. In addition, this data was already the result of generous beam time being provided for characterization (and is more than might be possible in many cases). In order to assess model performance when interpolating to new combinations of input settings we evaluated transfer learning from simulation in a case where the full operational range is present, to measurements in cases where the model needs to interpolate to new setting ranges. This could be an effective method for producing a surrogate model, particularly when only limited measured data is available.

\begin{figure}[t]
    \centering
    \includegraphics[width = 70mm]{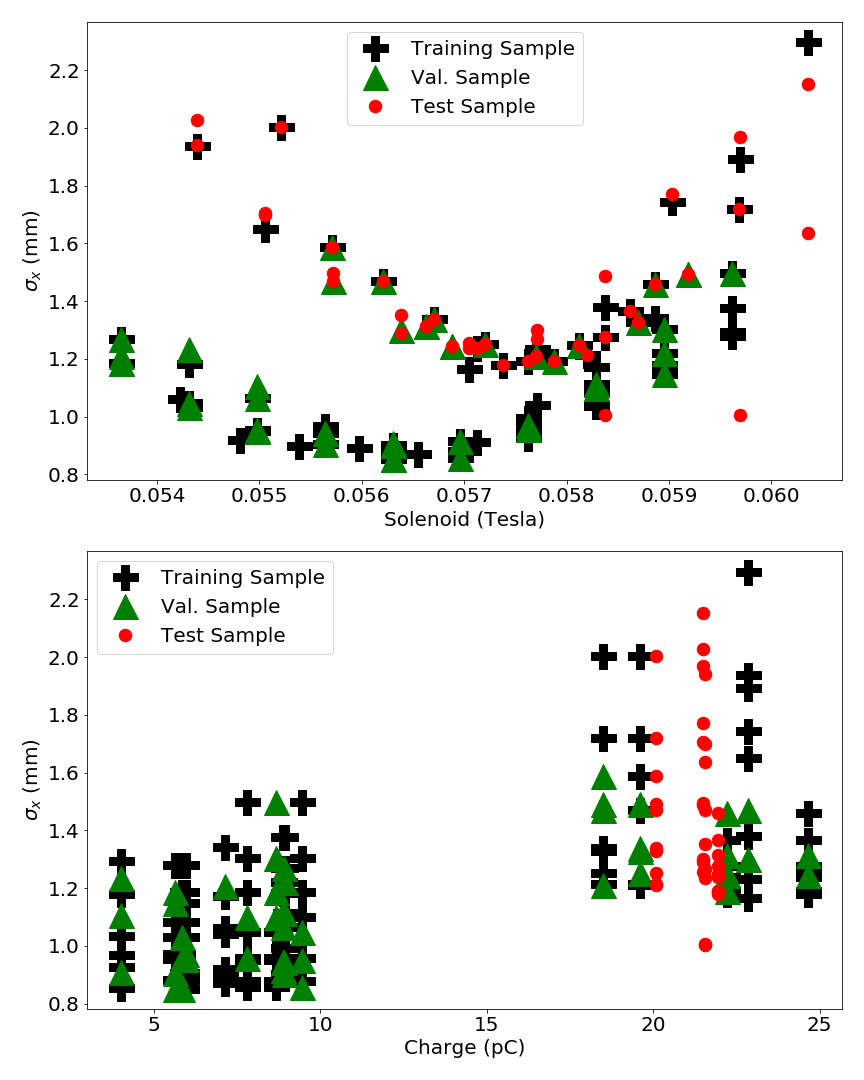}
    \caption{Training, validation, and testing samples for a surrogate model trained with only measured data, but with a large portion of measurements (in new beam charge ranges) withheld for test data. Shown are the 111 training samples, 48 validation samples, and 39 test samples.}
    \label{fig:sghm-trainingdata}
\end{figure}

This scenario (missing ranges of parameter space) was emulated by withholding measurement values with charge between 20~pC and 22~pC, with the data distribution shown in Fig.~\ref{fig:sghm-trainingdata}. The previously prototyped procedure was attempted, but we found we needed to adjust the  transfer learning procedure to accommodate the large systematic differences between the simulated and measure data. For the case with transfer learning to measured data, the main difference is the large systematic differences in the scalar output parameters, rather than in the types of input laser distributions seen during training. Thus, allowing more of the fully-connected layers to adapt to the new data was warranted. 

\begin{figure}[t]
    \centering
    \includegraphics[width=80mm]{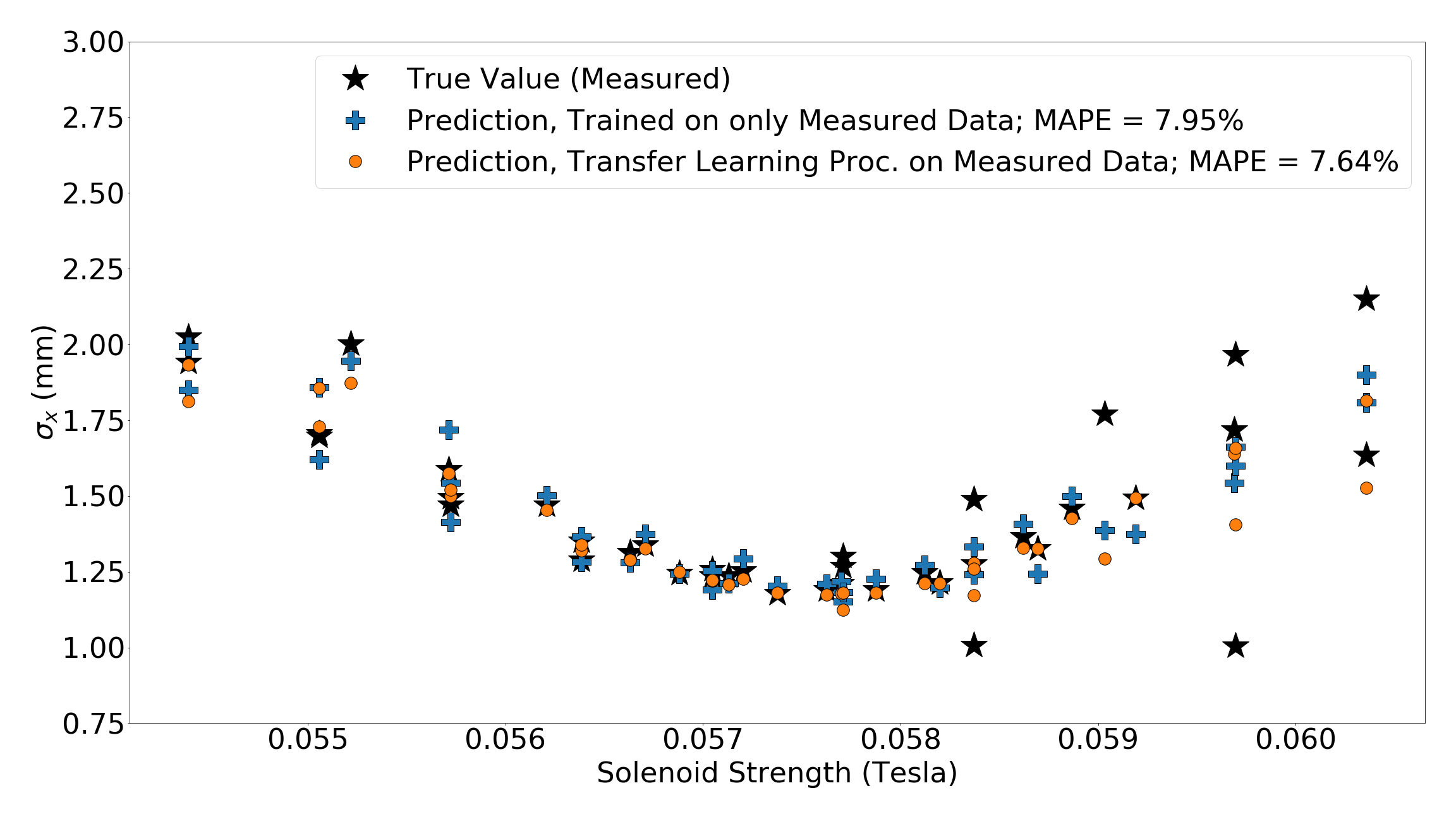}
    \caption{Prediction results for transfer learning between various models, predicting measured data. The base model was trained on VCC-based simulation data, whereas the measurement only data was trained using the data shown in Fig.~\ref{fig:h-traindata}. }
    \label{fig:sghm-solscan}
\end{figure}

\begin{figure}[t]
    \centering
    \includegraphics[width=80mm]{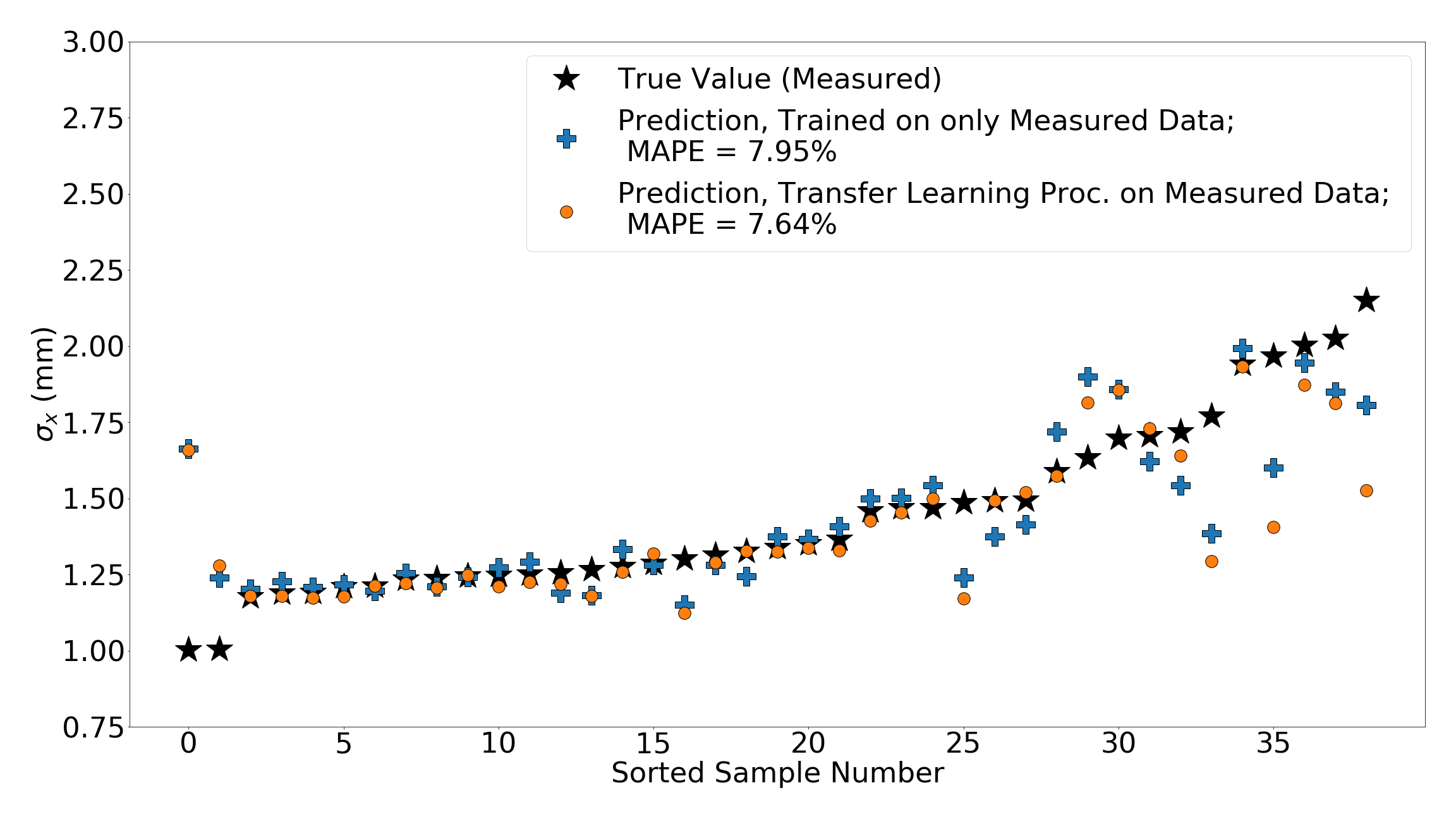}
    \caption{Prediction results for transfer learning between various models, predicting measured data. The base model was trained on VCC-based simulation data, whereas the measurement only data was trained using the data shown in Fig.~\ref{fig:h-traindata}. }
    \label{fig:sghm-predictions}
\end{figure}

The procedure was modified in the following ways. The base model (trained only on simulation data) for this procedure is the same as that produced previously during the simulation prototype. In this new case, the fully-connected portion (i.e. excluding the CNN layers of the neural network) were allowed to train with a reduced learning rate starting at \SI{5e-5} and decreasing every 10 epochs for 2000 epochs. The final learning rate is then used while annealing the model (with all layers trained) for another 2000 epochs. The results are shown in Fig. \ref{fig:sghm-predictions} and Fig. \ref{fig:sghm-solscan}


Here, the transfer learning procedure resulted in a model that can predict the measured data as well as the model trained only on measured data, at 7.95\% and 7.65\%. The MAPE on the original simulated data sets were similarly comparable, at 100\% and 119\% respectively. There is a clear trade off between the accuracy reached on the target data set (measured data) versus the base data set (simulation data). Iteration on this procedure may further improve the agreement on both data sets as needed for experimental use. Further, the transfer learning model is able to predict on a broader range of laser input distributions, is expected to generalize better to new beam distributions.

\section{Conclusions and Future Study}
Surrogate models are a viable solution for many challenges faced while designing and operating particle accelerators. They can be used for real-time feedback in the form of virtual diagnostics, for offline experiment planning, and many other applications. In our study, we demonstrated novel methods for designing and training more comprehensive injector surrogate models. 

First, a scalar surrogate model based on a wide range of simulated data was demonstrated, and we verified that it can be used for offline multi-objective optimization. Next, we showed that incorporating measured fluctuations in the initial laser distribution can improve surrogate model. Specifically, by including measured laser inputs during the training process, the model can more accurately predict beam outputs for out-of-distribution laser inputs. Previous injector surrogate models have not leveraged measured laser input fluctuations. Therefore, we showcase how important this inclusion is towards improving long term surrogate model viability during operation.  

Then, to train a simulation-based surrogate model trained on idealized laser distributions to be more representative of the real machine, a simple data augmentation technique and a transfer learning procedure was able to successfully learn both output distributions (ideal and VCC-generated). As the LCLS-II injector is operated, additional measured VCC images could be incorporated into the model using this approach. Other methods for data augmentation for improving disparity in sampling such as the Synthetic Minority Oversampling Technique (SMOTE) \cite{chawla2002smote} could also be tried and compared. 

Finally, we developed and applied a transfer learning procedure for transferring from simulation to measured data, which successfully reduced the model prediction error on a held out range of beam charges from 112.7\% to 7.6\%. Further iteration of the transfer learning process will likely improve the surrogate model training on both simulated and measured training data. 

Further, the simulated data can be expanded to include more operational ranges such as gun gradient values, which may help resolve the shift in beam waist seen in the measured data. The ability of the surrogate model to successfully interpolate predictions within the range of possible input parameters (demonstrated in this case for previously unseen charges) can be very helpful for quickly estimating output parameters without needing experimental data. Our study shows that this is possible with a comprehensive machine-learning based surrogate model for the LCLS-II injector frontend. While we demonstrate this only for the LCLS-II Injector frontend, these approaches for improving online modeling of injector systems could be easily adapted to other accelerator facilities.

\section{Acknowledgement}

The authors would like to acknowledge Jane Shtalenkova and Alex Saad for their invaluable help and expertise, and the METSD department at SLAC for CAD drawings. Funding for this work was provided in part by the U.S. Department of Energy, Office of Science, under Contracts No. DE-AC02-76SF00515 and No. DE-AC02-06CH11357, FWP 100494 through BES on B\&R code KC0406020. Funding for L. Gupta provided by the U.S. Department of Energy, Office of Science Graduate Student Research (SCGSR) Program, the U.S. National Science Foundation under Award No. PHY-1549132, the Center for Bright Beams, and the U. S. Dept. of Energy SCGSR, and the University of Chicago Department of Physics.

\bibliographystyle{abbrv}
\providecommand{\noopsort}[1]{}\providecommand{\singleletter}[1]{#1}%

\end{document}